\documentclass[sigconf,nonacm]{acmart}
\usepackage{fancyhdr}
\usepackage{xcolor}

\fancypagestyle{plain}{
}
\AtBeginDocument{%
  }

\setcopyright{none}
\renewcommand\footnotetextcopyrightpermission[1]{}
\acmConference[Preprint]{Preprint}{2026}{}
\acmDOI{}
\acmISBN{}

\begin{document}

\title{How Code Representation Shapes False-Positive Dynamics in Cross-Language LLM Vulnerability Detection}

\author{Maofei Chen}
\authornote{Corresponding author.}
\orcid{0009-0000-8083-4862}
\affiliation{%
	\institution{China Telecom Research Institute}
	\country{China}
}

\author{Laifu Wang}
\affiliation{%
	\institution{China Telecom Research Institute}
	\country{China}
}

\author{Yue Qin}
\affiliation{%
	\institution{China Telecom Research Institute}
	\country{China}
}

\author{Yuan Wang}
\affiliation{%
	\institution{China Telecom Research Institute}
	\country{China}
}

\author{Bo Wu}
\affiliation{%
	\institution{China Telecom Research Institute}
	\country{China}
}

\author{Dongxin Liu}
\affiliation{%
	\institution{China Telecom Research Institute}
	\country{China}
}

\renewcommand{\shortauthors}{Chen et al.}

\begin{abstract}
How code representation format shapes false positive behaviour in cross-language LLM vulnerability detection remains poorly understood. We systematically vary training intensity and code representation format---raw source text versus pruned Abstract Syntax Trees---at both training time and inference time, across two 8B-parameter LLMs (Qwen3-8B and Llama~3.1-8B-Instruct) fine-tuned on C/C++ data from the NIST Juliet Test Suite (v1.3) and evaluated on Java (OWASP Benchmark v1.2) and Python (BenchmarkPython v0.1).

Our central finding is that cross-language FPR reflects the joint effect of training-time and inference-time representation, rather than either one alone. Text fine-tuning drives FPR upward monotonically---Qwen3-8B escalates from 0.763 (zero-shot) through 0.866 (pilot) to 1.000 (full-scale)---while F1 remains stable (0.637--0.688), masking the collapse. Our results support \emph{surface-cue memorisation} as the primary explanation: text fine-tuning encodes C/C++-specific API names and syntactic idioms as vulnerability triggers that fire indiscriminately on target-language code. A \emph{cross-representation probe}---applying text-trained weights to AST-encoded input without retraining---isolates this mechanism: Qwen3-8B FPR drops from 0.866 to 0.583, and 37.2\% of false positives revert to true negatives under AST input alone. Counter-intuitively, direct AST fine-tuning does not preserve this benefit (FPR $\geq$ 0.970), as flat linearisation appears to introduce structural surface cues of its own. The pattern replicates across both model families, and on BenchmarkPython the AST probe yields FPR$=$0.554---within 2.9 percentage points of the Java result---despite maximal surface-syntax differences, substantially weakening a domain-shift explanation. These findings motivate a \emph{pre-deployment consistency gate}---running alerts through both text and AST paths---as a retraining-free filter for false-positive-sensitive settings, at the cost of reduced recall.
\end{abstract}

\begin{teaserfigure}
  \includegraphics[width=\textwidth]{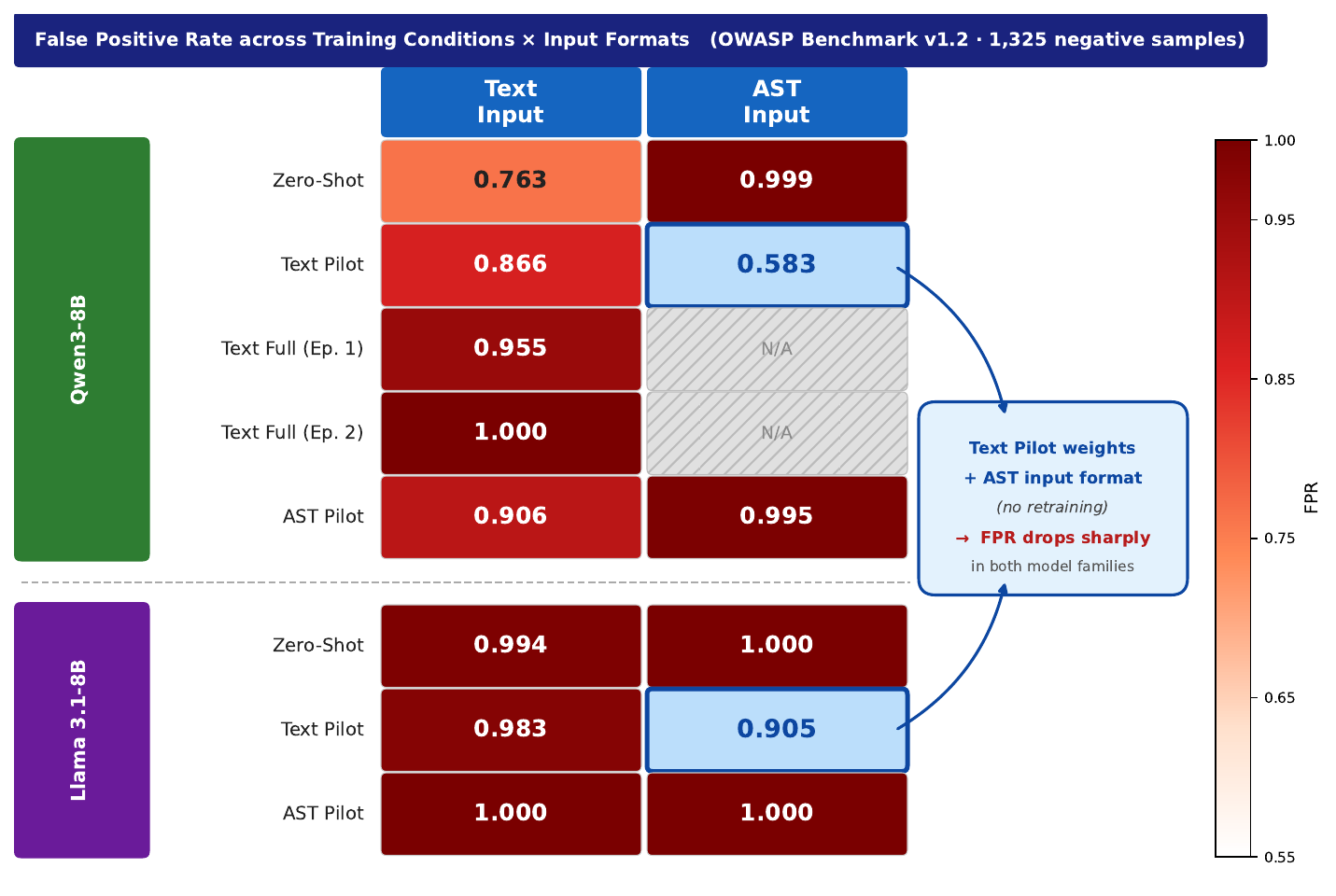}
  \caption{False Positive Rate (FPR) across training-condition $\times$ input-format
  combinations on the OWASP Benchmark v1.2 (1{,}325 Java negative samples),
  for Qwen3-8B and Llama~3.1-8B-Instruct fine-tuned on C/C++ (NIST Juliet v1.3)
  and evaluated on Java. Most cells show high FPR ($\geq$0.87); the
  highlighted ``Text Pilot $\to$ AST input'' cells are notable
  exceptions---applying text-trained weights to AST-encoded inputs, without
  retraining, reduces FPR in both families (Qwen: $0.866\to0.583$;
  Llama: $0.983\to0.905$), though absolute FPR remains high.
  See Table~\ref{tab:fpr-progression} for details.}
  \Description{Heatmap of False Positive Rates: most cells are deep red (FPR near 1.0),
    with two light-blue cells for Text Pilot weights applied to AST input showing
    substantially lower FPR (0.583 for Qwen3-8B, 0.905 for Llama 3.1-8B-Instruct).}
  \label{fig:teaser}
\end{teaserfigure}

\maketitle
\vspace{-1em}

\vspace{-0.5em}

\section{Introduction}
How do code representations at training time and inference time jointly shape false-positive behaviour in cross-language LLM vulnerability detection? This question has not been systematically characterised, yet it is operationally critical: high false positive rates remain a persistent challenge in LLM-based vulnerability detection~\cite{chendiversevul2023,gaohow2024,liuvuldetectbench2024}. We address it through a controlled characterisation study that independently varies both dimensions, identifying \emph{surface-cue memorisation} as the main explanatory account supported by our results and showing that FPR is a joint function of how a model is trained \emph{and} how it is queried---neither factor alone is sufficient.
In practice, the target language for vulnerability detection is determined by the codebase under analysis, whereas labelled training data is unevenly distributed across languages, making cross-language transfer a practical necessity rather than a deliberate choice. The standard workflow---fine-tune on available labelled data, ship when F1 looks acceptable---appears straightforward. Yet how training intensity and code representation format \emph{jointly} determine false positive behaviour in this cross-language setting has not been systematically characterised. This gap matters operationally: even state-of-the-art neural detectors consistently exhibit high false positive rates that limit their practical utility.
To address this gap, this paper presents a systematic characterisation study spanning two model families (Qwen3-8B and Llama~3.1-8B-Instruct), four training conditions, and two code representation formats (raw source text and pruned ASTs). By independently varying training-time and inference-time representations, we isolate their individual and joint effects on FPR. A key isolating experiment---the \emph{cross-representation probe}, applying text-trained weights to AST-encoded input without retraining---directly supports the surface-cue memorisation account. Testing the boundary condition (training on AST data) then corroborates the account and rules out simpler explanations: AST fine-tuning introduces its own memorisable structural cues, negating any representation advantage. All findings replicate across both model families. Our primary contributions are:
\begin{itemize}
    \item \textbf{First systematic characterisation of joint representation effects on FPR.} By independently varying training-time and inference-time code representations, we establish that cross-language FPR is governed by their \emph{interaction}, not by either factor alone. The cross-representation probe---text-trained weights applied to AST input without retraining---is the key isolating experiment: it yields a 32.6\% FPR reduction (Qwen3-8B: $0.866\to0.583$,~$\Delta{=}{-}0.282$; Llama~3.1-8B-Instruct: $0.983\to0.905$,~$\Delta{=}{-}0.078$) that directly supports the surface-cue account, not a deployable fix in isolation.

    \item \textbf{Surface-cue memorisation as the central explanatory account.} Text fine-tuning appears to encode C/C++-specific API names, string patterns, and syntactic idioms as vulnerability triggers; AST encoding recontextualises these lexical cues within a longer, more deeply nested representation, disrupting the source-form trigger patterns on Java code. At the case level, 427 of 1,147 Text Pilot false positives (37.2\%) are \emph{representation-sensitive}---they revert to correct negatives when only the input representation is switched---directly materialising the account as a countable, verifiable phenomenon. Per-Common Weakness Enumeration (CWE) heterogeneity corroborates the account: categories with purely lexical C/C++ signatures (e.g., CWE-330: $\Delta=-0.527$) improve far more than structurally prominent ones (e.g., CWE-78: $\Delta=-0.104$), precisely what lexical surface-cue dependence predicts. This account also explains why F1 remains stable across conditions even as FPR varies widely---F1 is an unreliable metric for cross-language SAST evaluation.

    \item \textbf{AST fine-tuning does not resolve the cross-representation effect.} Counter-intuitively, fine-tuning directly on AST-encoded data fails to preserve the FPR benefit: rates remain at or above 0.970 regardless of training format. Flat AST linearisation appears to introduce structural surface cues of its own, motivating richer semantic representations such as Program Dependence Graphs (PDGs) and Code Property Graphs (CPGs) as future work.

    \item \textbf{Replication across model families.} The above findings hold for both Qwen3-8B and Llama~3.1-8B-Instruct, two architecturally distinct model families, indicating that the effect is not a model-specific artefact but a general property of cross-language LLM deployment at the 8B scale.

    \item \textbf{Replication across target languages.} The pattern also replicates across target languages. On BenchmarkPython (v0.1, 1,108 Flask/Python samples), the AST probe achieves FPR$=$0.554---just \textbf{2.9pp} from the Java result (0.583)---despite maximal surface-syntax differences between the two target languages. This convergence provides the strongest evidence against a domain-shift explanation.
\end{itemize}
Taken together, this work provides the first systematic characterisation of how training intensity and code representation format jointly determine FPR dynamics in cross-language LLM-based vulnerability detection, with direct implications for the design of deployable SAST tools.

\section{Related Work}

\paragraph{LLMs for vulnerability detection.}
Transformer-based models pre-trained on code have become the dominant paradigm for automated vulnerability detection. CodeBERT~\cite{fengcodebert2020} established the bimodal pre-training approach, and LineVul~\cite{fulinevul2022} built on this to achieve state-of-the-art function- and line-level classification on C/C++ CVE data. DiverseVul~\cite{chendiversevul2023} showed that fine-tuned C/C++ models fail to generalise to unseen projects, with F1 dropping sharply on out-of-distribution code within the same language---a finding our work replicates and extends across the language boundary. Studies probing larger models via prompting~\cite{gaohow2024,liuvuldetectbench2024} find that even GPT-4-class models achieve limited detection performance on real-world code, suggesting the problem is scale-independent. MulVuln~\cite{nguyenmulvuln2025} and LLM4CVD~\cite{jianginvestigating2024} study multi-language pre-training and LLM model comparison, but neither systematically varies training intensity as an independent variable nor examines how input representation format modulates FPR in a cross-language deployment.

\paragraph{Cross-language and cross-domain transfer.}
CrossVul~\cite{nikitopouloscrossvul2021} provides a multi-language vulnerability dataset spanning 40+ languages; however, the work focuses on dataset construction rather than characterising the FPR dynamics induced by training transfer. Chakraborty et al.~\cite{chakrabortydeep2022} demonstrate that vulnerability detection models learn surface-level shortcuts rather than true vulnerability semantics, failing to generalise across datasets even within the same language---an insight that directly supports the surface-cue memorisation account we propose for the cross-language setting. The broader tendency of neural networks to exploit spurious statistical regularities---dubbed \emph{shortcut learning} by \citet{geirhosshortcut2020}---provides the theoretical grounding for this account: the model fits the shortest predictive path available in the training distribution, which in our setting is the co-occurrence of C/C++-specific surface tokens with vulnerability labels. Du et al.~\cite{dujoint2023} address cross-domain vulnerability detection via explicit geometrical and statistical distribution alignment. Concurrent work by Li et al.~\cite{liunveiling2024} studies project-specific shortcuts in same-language code models via token-level statistical measures (Cond-Idf), proposing a mitigation mechanism for intra-language bias. Our work differs in three respects: (1)~we examine cross-\emph{language} surface-cue memorisation rather than cross-project bias within a single language; (2)~we introduce inference-time representation switching as a zero-adaptation causal probe rather than a mitigation strategy requiring retraining; and (3)~we focus on FPR dynamics specifically, rather than general accuracy. Our work is complementary to Du et al.: we study the more prevalent case of standard supervised fine-tuning (SFT) \emph{without} domain adaptation and provide the first characterisation of the surface-cue memorisation account that explains FPR dynamics under cross-language deployment, showing that inference-time representation plays a decisive role in whether those memorised cues fire on target-language code.

\paragraph{Code representation for security analysis.}
Devign~\cite{zhoudevign2019} and SySeVR~\cite{lisysevr2022} demonstrated that rich code semantic representations---composite code graphs and AST/PDG-based program slices---outperform raw token sequences for in-language C/C++ detection. UniXcoder~\cite{guounixcoder2022} incorporated AST into multi-modal pre-training, improving general code representation quality. VuLASTE~\cite{zhuvulaste2023} extended AST encoding to long-sequence vulnerability detection. VulChecker~\cite{mirskyvulchecker2023} showed that fine-grained PDG representations achieve low FPR on C/C++ benchmarks. Our work contributes a complementary, cross-language finding: switching a text-fine-tuned model to AST-linearised input at inference time already reduces cross-language FPR substantially ($\Delta{=}{-}0.282$), whereas fine-tuning on AST data fails to sustain this benefit---confirming that flat AST linearisation is insufficient and richer data-flow representations (PDGs, CPGs) are needed.

\paragraph{Benchmarks and evaluation.}
The NIST Juliet Test Suite~\cite{bolandjuliet2012} provides balanced synthetic C/C++ programs widely used as a training corpus for ML-based vulnerability detectors~\cite{harerautomated2018}. Prior work established that such models tend to learn surface-level artefacts rather than true vulnerability semantics~\cite{chakrabortydeep2022}. Our work goes beyond confirming this phenomenon. To our knowledge, ours is the first to (a)~identify inference-time code representation as a controllable modulator of cross-language FPR, (b)~use representation switching as a zero-adaptation causal probe of the shortcut pattern without any retraining, and (c)~demonstrate cross-language FPR convergence---a 2.9pp gap between Java and Python under the same probe---as evidence that the pattern is language-invariant rather than a domain-shift artefact. The OWASP Benchmark v1.2~\cite{wichersowasp2015} provides 2,740 labelled Java test cases with an established TPR/FPR protocol; no prior work has evaluated a Juliet-fine-tuned LLM on OWASP, making this combination the first controlled study of its FPR dynamics.

\section{Methodology}


\subsection{Fine-tuning Dataset Preparation}
\label{sec:dataset-prep}

We constructed a leak-free SFT dataset from the NIST Juliet Test Suite for C/C++~\cite{bolandjuliet2012} through four stages.

\textbf{Scope and extraction.} Only intra-procedural test cases (source-to-sink logic within a single file) are included; inter-file variants are excluded. An adaptive extraction strategy handles two Juliet code patterns: for the majority of ``wrapper-worker'' cases, \textbf{logic inlining} merges the worker function body into the caller to produce a self-contained sample; for multi-call control-flow variants (e.g., \texttt{goodG2B1()}/\texttt{goodG2B2()}), each worker is treated as an independent sample to avoid conflating distinct logical scenarios.

\textbf{Data sanitization.} Three automated steps eliminate data leakage:
(1)~\textit{AST-based function anonymization}---CWE-revealing names
(e.g., \texttt{CWE15\_\allowbreak{}...\_bad}) are deterministically replaced with
neutral identifiers via \texttt{ast-grep};
(2)~\textit{comment removal}---all comments
(including \texttt{/* FLAW */}/\texttt{/* FIX */} hints) are stripped;
(3)~\textit{test harness elimination}---\texttt{main()} functions,
testing-specific directives, and non-essential headers are removed to isolate
vulnerability logic.

\textbf{AST representation optimization.} Semgrep-generated ASTs are pruned by removing OCaml serialization artifacts (\texttt{ref@} wrappers), transient semantic analysis data (\texttt{id\_resolved}, \texttt{id\_type}), and empty metadata containers, achieving a 79\% size reduction (from 19.4$\times$ to 4.0$\times$ source code length) while preserving all syntactic and structural information.

\textbf{Labeling.} Each sanitized snippet is formatted as a JSONL record: \texttt{\_bad} functions are labeled positive (\texttt{\{"found": true, "cwe": "CWE-XXX"\}}); \texttt{\_good} functions are labeled negative (\texttt{\{"found": false, "cwe": "N/A"\}}).

\textbf{Dataset splits.} We construct two training configurations. The \textbf{pilot} split contains only variant-01 test cases---the simplest, straight-line implementation of each vulnerability pattern without control-flow wrappers (${\sim}$3,100 training samples). The \textbf{full} split includes all Juliet variants (01--45), adding control-flow, data-flow, and platform wrappers around the same vulnerability cores (${\sim}$70,700 training samples). This two-tier design allows us to measure whether additional training data improves cross-language generalisation or merely deepens surface-cue memorisation. For the pilot split, text and AST training sets are effectively matched (3,100 vs.\ 3,099 samples). At full scale, 569 C/C++ samples (0.8\%) failed Semgrep AST conversion and were excluded from the AST training set, with most exclusions originating from CWE-415 good variants; the detailed per-CWE and per-label breakdown is reported in the replication package. CWE-415 is not among the evaluated OWASP categories.

\subsection{Prompt Engineering}
\label{sec:prompt}

To ensure that performance differences stem from model training rather than prompt engineering, all experiments use a unified system prompt template with only a minimal representation-specific preamble indicating whether the input is raw source text or a pruned AST; the task instructions and required JSON output format are otherwise identical. This standardization isolates the effect of structural representation (AST vs. text) from prompt formatting variations. The prompt requires the model to respond with only a JSON object containing two fields: \texttt{"found"} (boolean indicating vulnerability presence) and \texttt{"cwe"} (string with CWE identifier or ``N/A'').

The unified system prompt is defined as follows:
\begin{quote}
\small\raggedright
\texttt{You are an expert static analysis tool.\ Your task is to identify security vulnerabilities in the provided source code snippet.\ Your response MUST adhere strictly to the following JSON format.\ Do not include any reasoning, explanations, or text outside of the JSON object.}

\texttt{JSON format:}\\
\texttt{\{\ "found":\ <boolean>,\ "cwe":\ "<string>"\ \}}

\texttt{Instructions:\ (1)~"found":\ Use true if a vulnerability is found, otherwise false.\ (2)~"cwe":\ If "found" is true, provide the CWE ID (e.g., "CWE-89").\ If false, use "N/A".}
\end{quote}

All experiments reported in this paper---including all zero-shot baselines---use this prompt template exclusively.

\subsection{Fine-tuning Framework}
\label{sec:finetuning-framework}

We employ Low-Rank Adaptation (LoRA) for parameter-efficient fine-tuning of the Qwen3-8B model. The configuration uses a rank of $r=16$ and scaling factor $\alpha=32$, targeting the attention projection layers (\texttt{q\_proj}, \texttt{k\_proj}, \texttt{v\_proj}, \texttt{o\_proj}) in each transformer block. Training employs the AdamW optimizer with a learning rate of $2\times10^{-4}$, cosine decay scheduling with a warmup ratio of 0.02, and gradient checkpointing for memory efficiency. Hardware was allocated according to memory requirements across three GPU tiers: NVIDIA RTX 4090 (24\,GB), A100-PCIE (40\,GB), and A800 (80\,GB). The Qwen3-8B text pilot fit on the RTX 4090; all other pilot runs and full-scale text fine-tuning required the A100; full-scale AST fine-tuning used the A800 to accommodate the larger token sequences produced by AST serialisation. All experiments use mixed-precision (BF16) training.

\paragraph{Use of AI tools.} Large language models were used to assist in drafting and revising manuscript text for clarity and presentation. They were not used to generate experimental data, perform analyses, or make scientific decisions. All interpretations and the final manuscript text were reviewed and approved by the authors.

\section{Experiments and Results}
\subsection{Baseline Model Evaluation}
To establish a performance baseline, we evaluated the zero-shot capabilities of two code LLMs from different model families: Qwen3-8B and Llama~3.1-8B-Instruct. Both models were tested on the entire OWASP Benchmark v1.2 test suite, which comprises 2,740 Java code samples spanning 11 CWE categories, containing both vulnerable (true positive) and non-vulnerable (true negative) cases. The models' task was to identify potential vulnerabilities without any prior fine-tuning on a similar dataset.
All metrics are computed on binary vulnerability detection: a prediction is counted as positive when the model returns \texttt{"found": true}, regardless of the predicted CWE identifier. Per-CWE disaggregation uses the ground-truth CWE label from the OWASP expected-results file to partition samples, not the model's CWE prediction.

The overall performance metrics, aggregated across all vulnerability types, are presented below.

\begin{table}[htbp]
\centering
\small
\setlength{\tabcolsep}{3pt}
\caption{Overall Performance Metrics of the Baseline Models on the OWASP Benchmark v1.2 (N=2,740) across two input formats (raw source text and pruned AST). FPR is computed over 1,325 negative samples.}
\label{tab:baseline-performance}
\resizebox{\columnwidth}{!}{\begin{tabular}{llccccc}
\toprule
\textbf{Model} & \textbf{Input} & \textbf{F1} & \textbf{Recall} & \textbf{Prec.} & \textbf{Acc.} & \textbf{FPR} \\
\midrule
Qwen3-8B              & Text & 0.6526 & 0.8304 & 0.5375 & 0.5434 & 0.763 \\
Qwen3-8B              & AST  & 0.6813 & 1.0000 & 0.5166 & 0.5168 & 0.999 \\
Llama~3.1-8B-Instruct & Text & 0.6824 & 1.0000 & 0.5179 & 0.5193 & 0.994 \\
Llama~3.1-8B-Instruct & AST  & 0.6805 & 0.9986 & 0.5161 & 0.5157 & \textbf{1.000} \\
\bottomrule
\end{tabular}}
\end{table}

Both models exhibit the ``high-recall, low-precision'' pattern (Table~\ref{tab:baseline-performance}), but with different severity. Qwen3-8B retains residual discriminability (FPR$=$0.763), while Llama~3.1-8B-Instruct collapses nearly completely (FPR$=$0.994, only 8 true negatives across 1,325 safe samples). This convergence across two model families confirms that high FPR is a fundamental characteristic of current code LLMs applied to zero-shot vulnerability detection.

Zero-shot AST input uniformly collapses discriminability further: Qwen3-8B FPR rises from 0.763 to 0.999; Llama reaches 1.000. Only Qwen3-8B on raw text retains meaningful specificity; all other combinations converge to the ``flag everything'' regime (FPR$\geq$0.994). This confirms: \textbf{AST format alone provides no zero-shot benefit}. The FPR reduction in Section~\ref{sec:representation-impact} is a joint effect of fine-tuning \textit{combined with} AST format---neither alone is sufficient.

\subsection{Performance Analysis by Vulnerability Category}
\begin{figure}[htbp]
  \centering
  \includegraphics[width=\linewidth]{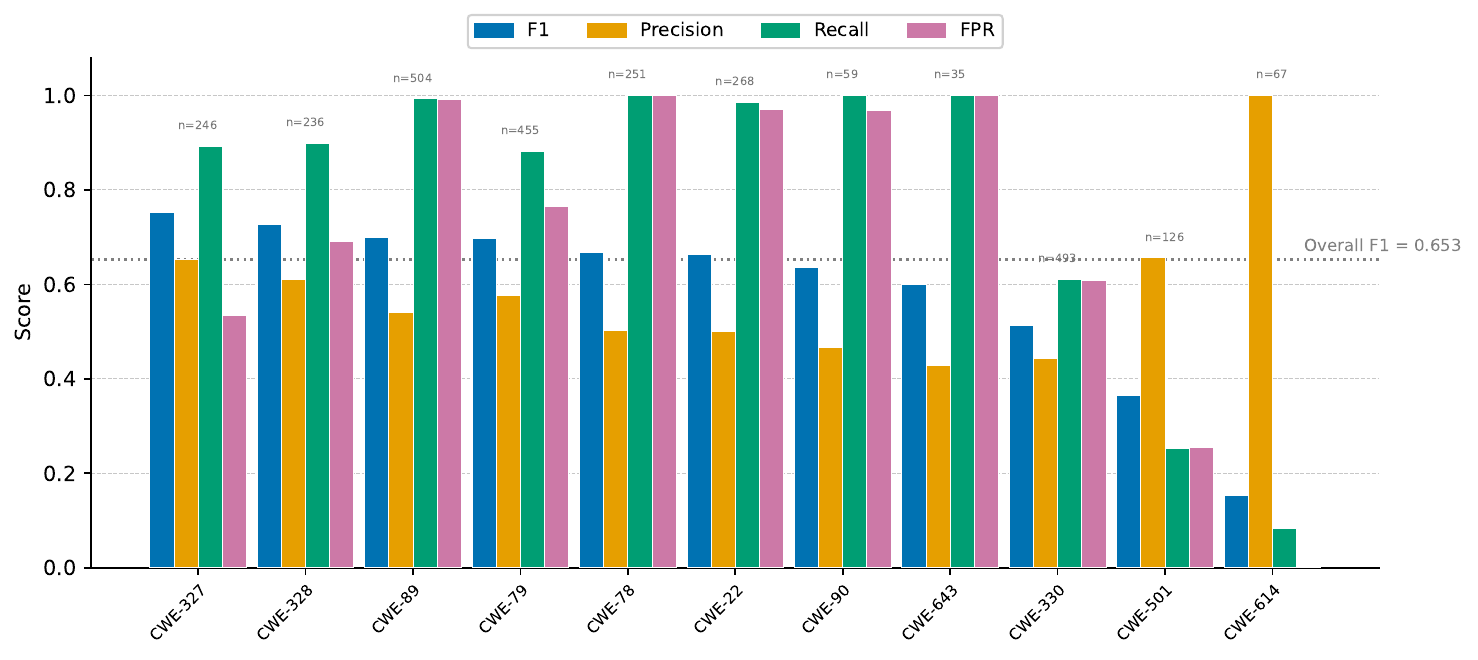}
  \caption{Per-CWE performance of the zero-shot Qwen3-8B model on the OWASP Benchmark (N=2,740).
           Categories are sorted by F1 score (descending).
           The dotted horizontal line marks the overall F1 of 0.653.
           Sample sizes ($n$) are shown above each bar group.}
  \label{fig:baseline-cwe}
\end{figure}

Figure~\ref{fig:baseline-cwe} disaggregates Qwen3-8B's zero-shot performance by CWE category (full confusion matrices in Appendix Table~\ref{tab:detailed-cwe-performance}).

The analysis reveals pronounced performance variance that foreshadows the surface-cue dynamics explored in later sections. Injection-type CWEs (CWE-89: FPR$=$0.991; CWE-78: FPR$=$1.000; CWE-22: FPR$=$0.970) achieve near-perfect recall, but this reflects indiscriminate triggering rather than selective pattern discrimination---the model labels nearly all samples as vulnerable regardless of ground truth, yielding high recall as a side effect. At the opposite extreme, CWE-614 (Insecure Cookie: Recall$=$0.083, FPR$=$0.000) exhibits near-total conservatism, predicting almost every sample as safe; CWE-501 (Trust Boundary Violation: Recall$=$0.253, FPR$=$0.256) follows a similar but less extreme pattern. These categories may have weaker or less consistently represented pretraining cues, leaving the model with little basis for confident positive prediction. CWE-330 (Weak Randomness: F1$=$0.514, FPR$=$0.607) occupies a middle ground: the model appears to possess partial lexical knowledge of weak-PRNG patterns (e.g., \texttt{Random} vs \texttt{SecureRandom}) from pretraining, but applies it imprecisely---a behaviour consistent with the high lexical dependence that the per-CWE analysis in Section~\ref{sec:representation-impact} later confirms ($\Delta$FPR$=$${-}$0.527, the largest improvement under AST input). This spectrum---from indiscriminate positive bias through partial lexical recognition to near-total conservatism---suggests that zero-shot vulnerability detection is shaped primarily by pre-existing lexical priors rather than semantic reasoning, with per-CWE behaviour determined by how strongly each vulnerability type is associated with recognisable surface patterns in the pretraining distribution.

\subsection{Fine-Tuning Results Overview}
\label{sec:finetune-results}

Table~\ref{tab:finetune-overall} presents the overall OWASP Benchmark performance across all experimental conditions for both model families. We label each condition descriptively: \textbf{Text Pilot} and \textbf{AST Pilot} refer to fine-tuning on the pilot split using source text or pruned AST, respectively; \textbf{Text Full} and \textbf{AST Full} refer to fine-tuning on the full split. For Full conditions, training was configured for three epochs; we report results from the epoch-1 and epoch-2 checkpoints, by which point training loss had largely plateaued. AST Full is reported at the epoch-2 checkpoint. Text Full reports epoch~1 and epoch~2 separately because the two checkpoints yield meaningfully different OWASP results (notably, FPR 0.955 vs.\ 1.000). Llama experiments are limited to the pilot split, as they serve to verify cross-family replication; training-scale effects are analysed with Qwen.

\begin{table}[htbp]
\centering
\small
\setlength{\tabcolsep}{3pt}
\caption{Overall OWASP Benchmark v1.2 Performance across all model--format combinations (N=2,740). The \textbf{OWASP} column indicates whether the benchmark samples were encoded as raw source text or pruned AST. FPR is computed over 1,325 negative samples.}
\label{tab:finetune-overall}
\resizebox{\columnwidth}{!}{\begin{tabular}{llccccc}
\toprule
\textbf{Condition} & \textbf{OWASP} & \textbf{F1} & \textbf{Recall} & \textbf{Prec.} & \textbf{Acc.} & \textbf{FPR} \\
\midrule
\multicolumn{7}{l}{\textit{Qwen3-8B: Zero-shot baseline}} \\
Zero-Shot               & Text & 0.6526 & 0.8304 & 0.5375 & 0.5434 & 0.763 \\
Zero-Shot               & AST  & 0.6813 & 1.0000 & 0.5166 & 0.5168 & 0.999 \\
\midrule
\multicolumn{7}{l}{\textit{Qwen3-8B: Text-based fine-tuning (C/C++ source text)}} \\
Text Pilot   & Text & 0.6875 & 0.9484 & 0.5392 & 0.5547 & 0.866 \\
Text Pilot   & AST  & 0.6372 & 0.7230 & 0.5696 & 0.5748 & 0.583 \\
Text Full (Ep.~1)                 & Text & 0.6726 & 0.9597 & 0.5177 & 0.5175 & 0.955 \\
Text Full (Ep.~2)                 & Text & 0.6811 & 1.0000 & 0.5164 & 0.5164 & 1.000 \\
\midrule
\multicolumn{7}{l}{\textit{Qwen3-8B: AST-based fine-tuning (pruned semgrep AST)}} \\
AST Pilot                & AST  & 0.6765 & 0.9873 & 0.5145 & 0.5124 & 0.995 \\
AST Pilot                & Text & 0.6800 & 0.9519 & 0.5289 & 0.5372 & 0.906 \\
AST Full (Ep.~2)               & AST  & 0.6791 & 0.9809 & 0.5193 & 0.5212 & 0.970 \\
AST Full (Ep.~2)               & Text & 0.6811 & 1.0000 & 0.5164 & 0.5164 & 1.000 \\
\midrule
\multicolumn{7}{l}{\textit{Llama~3.1-8B-Instruct: Zero-shot baseline}} \\
Zero-Shot                         & Text & 0.6824 & 1.0000 & 0.5179 & 0.5193 & 0.994 \\
Zero-Shot                         & AST  & 0.6805 & 0.9986 & 0.5161 & 0.5157 & 1.000 \\
\midrule
\multicolumn{7}{l}{\textit{Llama~3.1-8B-Instruct: Text-based fine-tuning (C/C++ source text)}} \\
Text Pilot   & Text & 0.6798 & 0.9887 & 0.5180 & 0.5190 & 0.983 \\
Text Pilot   & AST  & 0.6738 & 0.9385 & 0.5255 & 0.5307 & 0.905 \\
\midrule
\multicolumn{7}{l}{\textit{Llama~3.1-8B-Instruct: AST-based fine-tuning (pruned semgrep AST)}} \\
AST Pilot                & AST  & 0.6811 & 1.0000 & 0.5164 & 0.5164 & 1.000 \\
AST Pilot                & Text & 0.6811 & 1.0000 & 0.5164 & 0.5164 & 1.000 \\
\bottomrule
\end{tabular}}

\end{table}

Table~\ref{tab:finetune-overall} reveals four patterns, all replicating across both model families:
\textbf{(1)~Text on text: FPR escalates monotonically.} Qwen3-8B FPR climbs from 0.763 (zero-shot) to 0.866 (Text Pilot) to 1.000 (Text Full Ep.~2); Llama stays near ceiling ($0.994\to0.983$). F1 masks this collapse because perfect recall offsets catastrophic FPR.
\textbf{(2)~Zero-shot on AST: complete failure.} Both models default to vulnerability-prediction bias (Qwen FPR$=$0.999, Llama FPR$=$1.000)---AST format alone provides no benefit.
\textbf{(3)~Text Pilot on AST: FPR drops sharply.} Applying the \textit{same} text-trained weights to AST input without retraining reduces FPR substantially: Qwen $0.866\to$\textbf{0.583} ($\Delta=-0.282$), Llama $0.983\to$\textbf{0.905} ($\Delta=-0.078$). AST format appears to disrupt the triggers memorised during text training.
\textbf{(4)~AST Pilot on AST: FPR escalation resumes.} Qwen AST Pilot FPR$=$0.995; Llama$=$1.000---the model memorises C/C++ \textit{structural} patterns that fire indiscriminately on Java ASTs. Detailed analysis follows in Sections~\ref{sec:representation-impact} and~\ref{sec:cross-domain}.

\subsection{Cross-Domain Transfer Analysis}
\label{sec:cross-domain}

To gain a more nuanced understanding of the generalization capability, we disaggregate the results by CWE category and further partition them into two groups based on the overlap between our training corpus (NIST Juliet C/C++) and the evaluation benchmark (OWASP Java): \textbf{overlapping CWEs}, where the model was exposed to equivalent vulnerability patterns during fine-tuning (albeit in a different programming language), and \textbf{non-overlapping CWEs}, where the model receives no relevant training signal for that vulnerability type.

A systematic comparison between the CWE categories covered by the NIST Juliet C/C++ suite and those represented in the OWASP benchmark reveals that 6 of the 11 OWASP CWE categories (1,553 samples, 56.7\%) have corresponding entries in NIST, while the remaining 5 categories (1,187 samples, 43.3\%) are absent from the training data.

\begin{table*}[htbp]
\centering
\footnotesize
\setlength{\tabcolsep}{2pt}
\caption{Per-CWE Performance on OWASP Benchmark (Qwen3-8B). Each condition reports Recall (R), False Positive Rate (FPR), and F1; $\Delta$F1 is relative to zero-shot. Per-CWE FPR is computed over the negative samples within each CWE category.}
\label{tab:finetune-cwe}
\begin{tabular}{llr|ccc|cccc|cccc}
\toprule
& & & \multicolumn{3}{c|}{\textbf{Zero-Shot}} & \multicolumn{4}{c|}{\textbf{Text Pilot}} & \multicolumn{4}{c}{\textbf{AST Pilot}} \\
\textbf{CWE} & \textbf{Category} & \textbf{N} & \textbf{R} & \textbf{FPR} & \textbf{F1} & \textbf{R} & \textbf{FPR} & \textbf{F1} & \textbf{$\Delta$F1} & \textbf{R} & \textbf{FPR} & \textbf{F1} & \textbf{$\Delta$F1} \\
\midrule
\multicolumn{14}{l}{\textit{Overlapping CWEs (NIST training data available)}} \\
22  & pathtraver     & 268 & .985 & .970 & .663 & .985 & .933 & .672 & $+$.009 & .970 & 1.00 & .650 & $-$.013 \\
78  & cmdi           & 251 & 1.00 & 1.00 & .668 & 1.00 & .992 & .670 & $+$.002 & 1.00 & 1.00 & .668 &    .000 \\
90  & ldapi          &  59 & 1.00 & .969 & .635 & .889 & .750 & .640 & $+$.005 & 1.00 & .969 & .635 &    .000 \\
327 & crypto         & 246 & .892 & .534 & .753 & .808 & .603 & .689 & $-$.065 & .931 & .974 & .665 & $-$.088 \\
328 & hash           & 236 & .899 & .692 & .727 & .961 & .944 & .701 & $-$.027 & .992 & 1.00 & .703 & $-$.024 \\
330 & weakrand       & 493 & .610 & .607 & .514 & .945 & .771 & .648 & $+$.134 & 1.00 & 1.00 & .613 & $+$.100 \\
\midrule
\multicolumn{14}{l}{\textit{Non-Overlapping CWEs (no NIST training data)}} \\
89  & sqli           & 504 & .993 & .991 & .700 & .993 & .935 & .712 & $+$.012 & .985 & .987 & .697 & $-$.002 \\
79  & xss            & 455 & .882 & .766 & .697 & .931 & .909 & .689 & $-$.008 & 1.00 & 1.00 & .702 & $+$.005 \\
501 & trustbound     & 126 & .253 & .256 & .365 & .928 & .884 & .778 & $+$.413 & 1.00 & 1.00 & .794 & $+$.429 \\
614 & securecookie   &  67 & .083 & .000 & .154 & .972 & .807 & .729 & $+$.575 & 1.00 & 1.00 & .699 & $+$.545 \\
643 & xpathi         &  35 & 1.00 & 1.00 & .600 & 1.00 & 1.00 & .600 &    .000 & 1.00 & 1.00 & .600 &    .000 \\
\midrule
\textbf{All} & & \textbf{2740} & & & \textbf{.653} & & & \textbf{.688} & \textbf{$+$.035} & & & \textbf{.677} & \textbf{$+$.024} \\
\bottomrule
\end{tabular}
\end{table*}

\begin{figure*}[htbp]
  \centering
  \includegraphics[width=\linewidth]{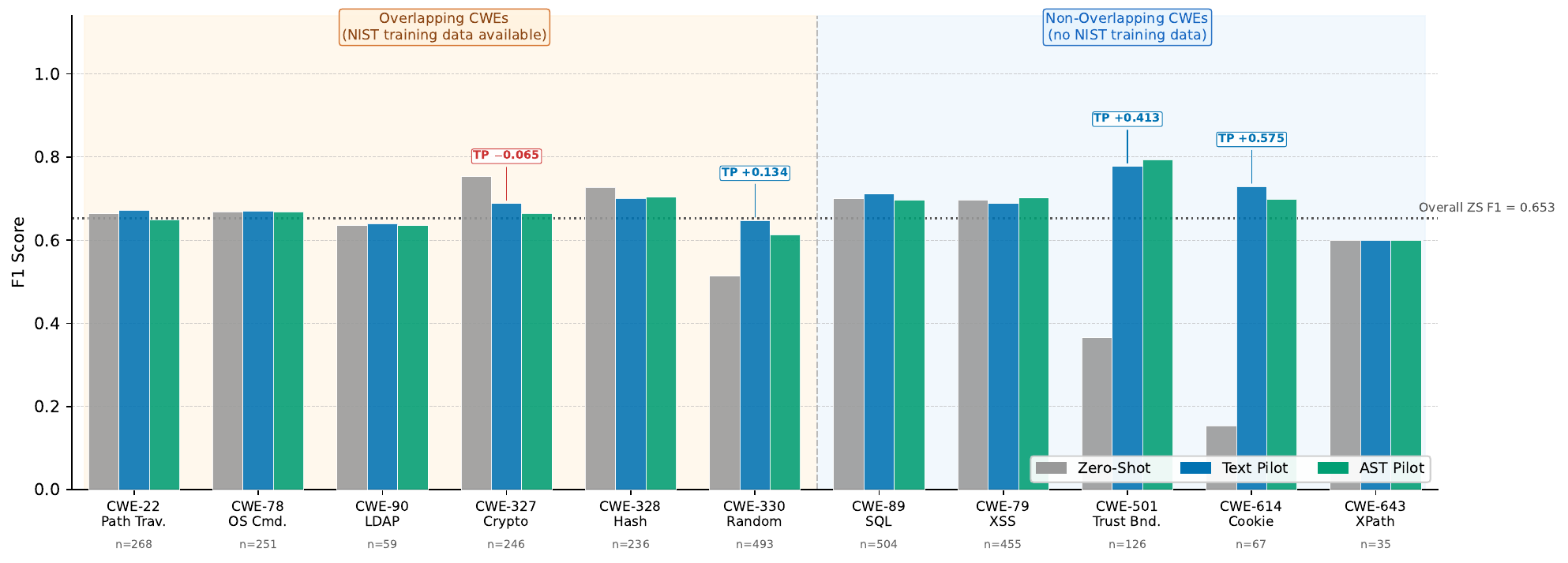}
  \caption{Per-CWE F1 score comparison on the OWASP Benchmark v1.2 (N=2,740) for
    Zero-Shot, Text Pilot, and AST Pilot (Qwen3-8B).
    CWEs are grouped by overlap with the NIST C/C++ training corpus.
    \textbf{Overlapping CWEs} (left, warm background): F1 changes are modest;
    CWE-330 shows the largest F1 increase ($+$0.134), though concurrent FPR rise tempers
    the improvement (Table~\ref{tab:finetune-cwe}).
    \textbf{Non-Overlapping CWEs} (right, blue background): CWE-501 ($+$0.413) and CWE-614
    ($+$0.575) show large F1 gains, but these reflect a conservative-to-aggressive shift in
    prediction behaviour rather than uniformly improved discrimination (see Recall/FPR
    decomposition in Table~\ref{tab:finetune-cwe}).
    The dotted line marks the overall zero-shot F1 baseline (0.653).}
  \label{fig:finetune-cwe}
\end{figure*}

Our earlier critique of F1 concerns aggregate evaluation, where rising recall can mask catastrophic false-positive growth. In this per-CWE analysis, F1 is retained only as a within-category summary; interpretation relies primarily on the accompanying Recall and FPR decomposition.

The per-CWE results (Table~\ref{tab:finetune-cwe} and Figure~\ref{fig:finetune-cwe}) reveal three distinct patterns when Recall and FPR are examined alongside F1.

\textbf{(1)~Conservative-to-aggressive shift on under-detected CWEs.} CWE-614 and CWE-501---non-overlapping categories absent from training data---show the largest F1 gains ($+$0.575 and $+$0.413, respectively), but Recall/FPR decomposition reveals that these gains are driven by a wholesale shift in prediction behaviour rather than improved discrimination. CWE-614 moves from extreme conservatism (R$=$0.083, FPR$=$0.000) to near-universal positive prediction (R$=$0.972, FPR$=$0.807); CWE-501 follows a similar trajectory (R: $0.253\to0.928$, FPR: $0.256\to0.884$). Fine-tuning recalibrates the model toward emitting positive vulnerability judgements in categories that zero-shot treated conservatively, but this shift comes with a sharp loss of specificity. Since these CWEs are absent from training data, the explanation cannot be domain knowledge transfer; rather, fine-tuning appears to lower the model's overall decision threshold. This threshold shift provides indirect support for the surface-cue memorisation account: even where no category-specific transfer is possible, fine-tuning appears to increase the model's overall propensity to emit ``vulnerable'' judgements, consistent with an accumulation of C/C++-specific surface triggers that raises the aggregate positive-prediction rate.

\textbf{(2)~Partial cross-language transfer with concurrent FPR increase.} Among overlapping CWEs, CWE-330 (Weak Randomness) shows Recall improving from 0.610 to 0.945, potentially reflecting partial transfer of weak-PRNG detection patterns, though the general threshold shift from Pattern~(1) may also contribute given the sparse pilot-set representation (6 training samples). FPR rises concurrently ($0.607\to0.771$), confirming that the gain reflects increased sensitivity rather than a clean improvement in discriminative ability.

\textbf{(3)~FPR deterioration masked by stable F1.} CWE-328 exemplifies how per-CWE F1 can also obscure operational degradation: F1 changes by only $-$0.027, but FPR escalates from 0.692 to 0.944 as Recall rises from 0.899 to 0.961---the same masking dynamic observed at the aggregate level. Similarly, CWE-327 sees both Recall decline ($0.892\to0.808$) and FPR rise ($0.534\to0.603$), a double deterioration likely exacerbated by the extremely sparse pilot-set representation (6 training samples each for CWE-327 and CWE-328).

Across these representative patterns, per-CWE FPR typically remains elevated or worsens after fine-tuning, reinforcing the overall finding that text-based fine-tuning consistently increases false-positive propensity regardless of whether the CWE category is represented in the training data.

\subsection{Impact of Code Representation}
\label{sec:representation-impact}

\begin{figure}[htbp]
  \centering
  \includegraphics[width=\linewidth]{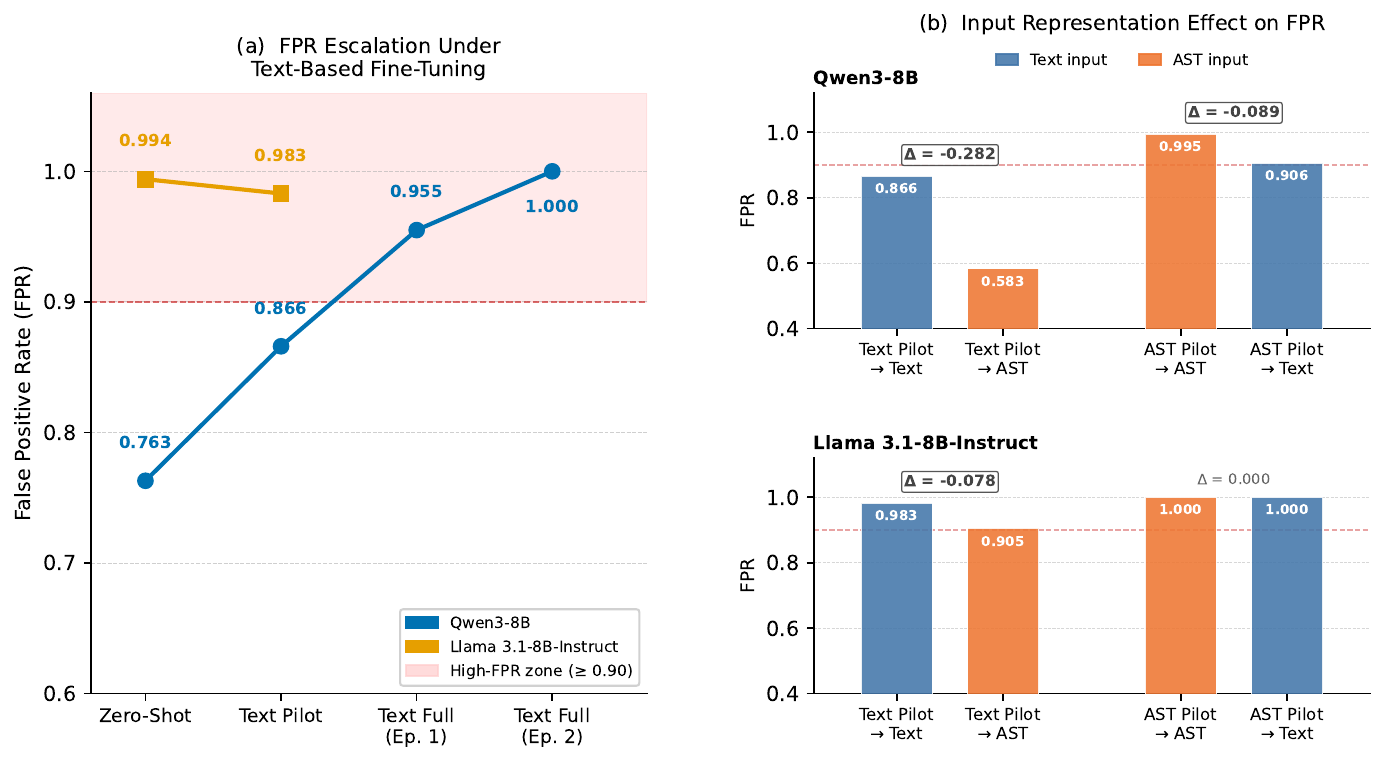}
  \caption{FPR progression across training conditions for Qwen3-8B and Llama~3.1-8B-Instruct.
    \textbf{(a)}~Under text-based fine-tuning (text-format evaluation), Qwen3-8B's FPR escalates
    monotonically from 0.763 (zero-shot) to 1.000 (Text Full, Ep.~2); Llama~3.1-8B-Instruct
    already operates near the ceiling at 0.994 zero-shot (Ep.~1--2 not run).
    Shaded region marks FPR$\geq$0.90.
    \textbf{(b)}~Input representation effect on FPR.
    The forward probe (Text Pilot$\to$AST) substantially reduces FPR for both models
    (Qwen: $0.866\!\to\!0.583$, $\Delta\!=\!{-0.282}$;
     Llama: $0.983\!\to\!0.905$, $\Delta\!=\!{-0.078}$),
    whereas the reverse probe (AST Pilot$\to$Text) is much weaker
    (Qwen: $0.995\!\to\!0.906$, $\Delta\!=\!{-0.089}$)
    or null (Llama: $1.000\!\to\!1.000$, $\Delta\!=\!0.000$).
    This asymmetry indicates that the low-FPR effect is a joint consequence
    of text-based fine-tuning and AST input, not an intrinsic property
    of either AST training or AST input alone.}
  \label{fig:fpr-progression}
\end{figure}

\textbf{FPR escalation as a fundamental failure mode of text-based fine-tuning.}
Qwen3-8B's zero-shot FPR of 0.763 already flags 76.3\% of safe Java samples. Text fine-tuning on NIST C/C++ data amplifies this systematically: Table~\ref{tab:fpr-progression} and Figure~\ref{fig:fpr-progression}(a) show FPR escalating from 0.763 to 0.866 (Text Pilot) to 1.000 (Text Full Ep.~2). The model appears to memorize C/C++-specific surface patterns---function names, API calls, syntactic idioms---that activate indiscriminately on safe Java code, driving FPR toward 1.0. This degradation is masked by F1 (Ep.~2: 0.681 vs.\ zero-shot 0.653) because perfect recall offsets catastrophic FPR, illustrating why \textbf{FPR is the operationally critical metric}.
The directional nature of this degradation is itself mechanistically informative: FPR rises sharply while Recall improves and F1 remains stable. This pattern is less consistent with a generic domain-shift explanation, which would more naturally produce broader degradation, and is more consistent with shortcut learning~\cite{geirhosshortcut2020}: the classifier becomes increasingly aggressive in predicting vulnerabilities whenever inputs resemble memorised C/C++ surface patterns. Corroborating this, 93.1\% of zero-shot false positives persist after text fine-tuning---fine-tuning \emph{reinforces} pre-existing biases rather than introducing new error modes.

\begin{table}[htbp]
\centering
\small
\setlength{\tabcolsep}{3.5pt}
\caption{FPR, Recall, and F1 across all training conditions and input formats on OWASP Benchmark v1.2 (1,325 negative samples). Rows are grouped by training condition so that each cross-representation probe (\textit{italics}) appears directly below its native-format baseline. $\Delta$FPR is defined only for italicised probe rows and is computed relative to the baseline immediately above, i.e., the change in FPR when switching input format without retraining. The format-mismatch effect is asymmetric: switching a text-trained model to AST input reduces FPR substantially (Qwen: $-0.282$; Llama: $-0.078$), whereas switching an AST-trained model to text input yields a much weaker and less consistent effect. Text Full rows show the FPR escalation trajectory under increased training intensity and have no cross-representation pair.}
\label{tab:fpr-progression}
\resizebox{\columnwidth}{!}{%
\begin{tabular}{llcccc}
\toprule
\textbf{Condition} & \textbf{Input} & \textbf{FPR} & \textbf{Recall} & \textbf{F1} & \textbf{$\Delta$FPR} \\
\midrule
\multicolumn{6}{l}{\textit{Qwen3-8B}} \\[2pt]
Zero-Shot                              & Text & 0.763 & 0.830 & 0.653 & --- \\
Zero-Shot                              & AST  & 0.999 & 1.000 & 0.681 & --- \\[2pt]
Text Pilot                             & Text & 0.866 & 0.948 & 0.688 & --- \\
\textit{AST (no retrain)} & AST  & \textbf{0.583} & 0.723 & 0.637 & $\mathbf{-0.282}$ \\[2pt]
AST Pilot                              & AST  & 0.995 & 0.987 & 0.677 & --- \\
\textit{Text (no retrain)} & Text & 0.906 & 0.952 & 0.680 & $-0.089$ \\[2pt]
AST Full (Ep.~2)                       & AST  & 0.970 & 0.981 & 0.679 & --- \\
\textit{Text (no retrain)} & Text & 1.000 & 1.000 & 0.681 & $+0.030$ \\[2pt]
\multicolumn{6}{l}{\quad\textit{--- Text Full trajectory (no cross-repr pair) ---}} \\
Text Full (Ep.~1)                      & Text & 0.955 & 0.960 & 0.673 & --- \\
Text Full (Ep.~2)                      & Text & 1.000 & 1.000 & 0.681 & --- \\
\midrule
\multicolumn{6}{l}{\textit{Llama~3.1-8B-Instruct}} \\[2pt]
Zero-Shot                              & Text & 0.994 & 1.000 & 0.682 & --- \\
Zero-Shot                              & AST  & 1.000 & 0.999 & 0.681 & --- \\[2pt]
Text Pilot                             & Text & 0.983 & 0.989 & 0.680 & --- \\
\textit{AST (no retrain)} & AST  & \textbf{0.905} & 0.939 & 0.674 & $\mathbf{-0.078}$ \\[2pt]
AST Pilot                              & AST  & 1.000 & 1.000 & 0.681 & --- \\
\textit{Text (no retrain)} & Text & 1.000 & 1.000 & 0.681 & $0.000$ \\
\bottomrule
\end{tabular}}
\end{table}

\textbf{Cross-representation test: AST breaks the FPR escalation pattern.}
To isolate the effect of input representation, we conducted a \textbf{cross-representation test}: we applied the Text Pilot model—trained exclusively on raw C/C++ source code—to OWASP Java samples encoded as pruned semgrep Generic ASTs, without any adapter retraining. This zero-adaptation probe yields a striking and practically significant finding.

\begin{table}[htbp]
\centering
\small
\setlength{\tabcolsep}{3pt}
\caption{Cross-Representation Test: Text Pilot Model on OWASP Benchmark v1.2 (N=2,740). Same model weights applied to two input formats. Negative sample count: 1,325.}
\label{tab:cross-repr}
\resizebox{\columnwidth}{!}{\begin{tabular}{lccccc}
\toprule
\textbf{Input Format} & \textbf{F1} & \textbf{Recall} & \textbf{Precision} & \textbf{FPR} & \textbf{FP count} \\
\midrule
Raw Source Text (in-distribution)  & 0.6875 & 0.9484 & 0.5392 & 0.866 & 1,147 \\
Pruned AST (out-of-distribution)   & 0.6372 & 0.7230 & 0.5696 & 0.583 &   773 \\
\midrule
$\Delta$ (AST $-$ Text) & $-$0.050 & $-$0.225 & $+$0.030 & $\mathbf{-0.282}$ & $\mathbf{-374}$ \\
\bottomrule
\end{tabular}}

\end{table}

\textbf{Key finding: prediction outcomes are causally affected by input representation.} The $2\times2$ pilot comparison (Table~\ref{tab:fpr-progression}) indicates that the effect of input representation depends on the model's fine-tuning trajectory, rather than reflecting a uniform benefit of AST format itself. For Qwen3-8B, switching the Text Pilot model from text to AST input reduces FPR substantially ($0.866\to0.583$, $\Delta=-0.282$; McNemar $p \ll 10^{-20}$), whereas AST input in zero-shot does not help ($0.763\to0.999$), ruling out a simple format-only explanation. At the case level, 427 of 1,147 false positives (37.2\%; Table~\ref{tab:cross-repr}) revert to true negatives under AST input with the same model checkpoint, providing concrete evidence that, for a substantial subset of errors, prediction outcomes are causally affected by input representation alone. However, this effect is asymmetric across models and training stages, and should therefore be interpreted as a representation-by-training interaction rather than a general advantage of AST input. The reduction in FPR is accompanied by only a moderate drop in F1 ($0.688\to0.637$, $\Delta=-0.05$), but also by a substantial decrease in recall ($0.948\to0.723$). This recall cost limits the probe's utility as a standalone filter, but does not undermine its diagnostic value for identifying representation-sensitive errors.

\textbf{Interpretation: AST input appears to disrupt memorised C/C++ surface triggers.} The text-trained model's predictions are strongly shaped by surface-level textual cues---such as API names, string patterns, and syntactic idioms---consistent with shortcut learning~\cite{geirhosshortcut2020}. In the OWASP Java corpus, safe code routinely invokes APIs that resemble vulnerability-associated patterns (e.g., cryptographic functions, path-manipulation utilities), contributing to indiscriminate firing on text input (FPR$=$0.866). AST encoding recontextualises these lexical cues---embedding them within a longer, more deeply nested representation that disrupts the source-form trigger templates learned during text fine-tuning---reducing this false-alarm pattern (FPR$=$0.583).

Critically, this reduction is \textit{not} an intrinsic property of AST format. The zero-shot model given the same AST input achieves FPR$=$0.999, indicating that AST input alone is not beneficial. Rather, without task-specific adaptation to this representation, the model appears to treat AST inputs as uninformative or out-of-distribution and responds with near-universal positive predictions. The observed reduction is therefore best understood as a representation-by-training interaction: the model must be sufficiently fine-tuned to produce calibrated task outputs, while the AST input suppresses lexical anchors that would otherwise trigger many false positives under text input. Neither condition alone is sufficient.

\textbf{Asymmetric format-mismatch effect.}
The full cross-representation matrix (Table~\ref{tab:fpr-progression}) reveals a telling asymmetry. Training on AST data does \textit{not} fix the FPR problem on AST input (FPR$=$0.995); however, applying the AST Pilot model to text input yields FPR$=$0.906 ($\Delta=-0.089$), while the reverse direction (Text Pilot on AST input) produces a much larger $\Delta=-0.282$. This \textbf{asymmetry is informative about representation dependence}:

\begin{itemize}
  \item \textbf{Text model $\xrightarrow{}$ AST input (strong effect, $\Delta=-0.282$):} The text model's false-positive triggers appear to be strongly format-dependent: AST serialisation disrupts the source-form patterns (API names, string literals, syntactic idioms) on which these triggers rely, reducing their activation on safe samples.
  \item \textbf{AST model $\xrightarrow{}$ Text input (weak effect, $\Delta=-0.089$):} One possibility is that raw source text exposes surface proxies for structural patterns (method signatures, code layout) that remain sufficiently aligned with AST-learned cues, partially sustaining false-positive triggers. A more fundamental consideration is \emph{pretraining modality bias}: both base models are pretrained predominantly on raw source text, so text is closer to the model's native input distribution than AST. LoRA fine-tuning adapts only a thin parameter subspace (rank$=$16, targeting attention projections), which may be insufficient to override the base model's text-conditioned priors. Under this account, the AST$\to$text mismatch appears weak not because the AST model ``partially activates AST triggers on text,'' but because the underlying text-native representations were never fully overwritten. Additionally, a \emph{ceiling effect} may compress the observable $\Delta$: the AST Pilot already exhibits near-saturated FPR on matched input (0.995), leaving little room for further increase under text input. These explanations are not mutually exclusive, and the present matrix does not distinguish among them.
\end{itemize}

\textbf{Cross-family replication.}
Before turning to per-CWE analysis, we check whether the global patterns above hold across model families. The same four qualitative signatures also appear for Llama~3.1-8B-Instruct (Table~\ref{tab:fpr-progression}): Text Pilot remains near the FPR ceiling ($0.994\to0.983$), zero-shot AST reaches ceiling (FPR$=$1.000), the cross-representation probe still reduces FPR to 0.905 ($\Delta=-0.078$), and AST Pilot remains saturated on both formats (FPR$=$1.000). The smaller observed $\Delta$ for Llama is plausibly explained in part by a \textbf{ceiling effect}, since its matched text-condition FPR already lies in the near-saturated range, compressing the observable reduction under the FPR$\leq$1.0 upper bound. This smaller magnitude should therefore be interpreted cautiously. Overall, the convergence across two distinct model families suggests that the main patterns identified here are not idiosyncratic to a single model, at least within this benchmark setting.

\textbf{Per-CWE heterogeneity as mechanistic evidence.}
Having established the global picture and its cross-family robustness, we now examine how the cross-representation effect distributes across individual vulnerability types. Table~\ref{tab:cwe-lexical-tiers} ranks all 11~CWEs by $\Delta$\,=\,FPR(AST)\allowbreak$\,-\,$FPR(Text), revealing a broad gradient rather than a uniform shift; this pattern is consistent with the surface-cue account and difficult to reconcile with a uniform ``AST is harder'' explanation (full matrix in Appendix~\ref{appendix:cwe-fpr}). The remaining 720 persistent false positives show that a substantial portion of the error burden is not removed by changing representation alone, suggesting additional representation-invariant factors. This per-CWE gradient is clearest for Qwen3-8B; Llama~3.1-8B shows qualitatively similar trends but with compressed magnitudes (Appendix~\ref{appendix:cwe-fpr}).

\begin{table}[htbp]
\centering
\footnotesize
\setlength{\tabcolsep}{4pt}
\caption{Per-CWE FPR under the cross-representation probe (Text Pilot weights, OWASP Java, 1,325 negatives), sorted by $\Delta$ = FPR(AST) $-$ FPR(Text) within tier. Tier labels: \textbf{H}igh / \textbf{M}edium / \textbf{L}ow lexical dependence; $\oplus$ = exception (FPR rises under AST).}
\label{tab:cwe-lexical-tiers}
\begin{tabular}{llccc}
\toprule
\textbf{CWE} & \textbf{Category} & \textbf{FPR\textsubscript{text}} & \textbf{FPR\textsubscript{AST}} & \textbf{$\Delta$} \\
\midrule
\multicolumn{5}{l}{\textit{(H) High lexical dependence\ \ ($\Delta \leq -0.3$)}} \\
330 & Insuff.\ random.   & 0.771 & 0.244 & $-$0.527 \\
643 & XPath injection    & 1.000 & 0.550 & $-$0.450 \\
614 & Sensitive cookie   & 0.806 & 0.387 & $-$0.419 \\
22  & Path traversal     & 0.933 & 0.526 & $-$0.407 \\
79  & XSS                & 0.909 & 0.531 & $-$0.378 \\
\midrule
\multicolumn{5}{l}{\textit{(M) Medium\ \ ($-0.3 < \Delta \leq -0.1$)}} \\
328 & Weak hash          & 0.944 & 0.682 & $-$0.262 \\
501 & Trust boundary     & 0.884 & 0.651 & $-$0.233 \\
89  & SQL injection      & 0.935 & 0.797 & $-$0.138 \\
78  & Command inject.    & 0.992 & 0.888 & $-$0.104 \\
\midrule
\multicolumn{5}{l}{\textit{(L) Low\ \ ($|\Delta| < 0.1$)}} \\
90  & LDAP injection     & 0.750 & 0.688 & $-$0.062 \\
\midrule
\multicolumn{5}{l}{\textit{($\oplus$) Exception\ \ ($\Delta > 0$)}} \\
327 & Weak crypto alg.   & 0.603 & 0.707 & $+$0.103 \\
\midrule
\textbf{All} & & \textbf{0.866} & \textbf{0.583} & $\mathbf{-}$\textbf{0.282} \\
\bottomrule
\end{tabular}
\end{table}

For descriptive clarity, we group CWEs into high-, medium-, and low-shift tiers by~$\Delta$. The five \textbf{H}-tier CWEs ($\Delta \leq -0.3$) are those whose vulnerability signatures are most strongly associated with distinctive API names and string literals, whose source-level trigger patterns appear to be most disrupted under AST encoding. CWE-330 provides the sharpest example ($\Delta=-0.527$): although identifiers such as \texttt{Random} and \texttt{SecureRandom} remain visible in the semgrep\_generic AST, AST encoding recontextualises them within a longer, more deeply nested representation, disrupting the source-form trigger templates learned during text fine-tuning. The four \textbf{M}-tier CWEs ($-0.3 < \Delta \leq -0.1$) involve structurally salient call-site patterns (e.g., SQL construction, command execution) that remain partially visible after AST encoding---the source-level lexical trigger pattern is weakened but the structural trigger persists. CWE-90 (\textbf{L} tier, $\Delta=-0.062$) is closer to a data-flow pattern already represented at the AST level, leaving less opportunity for a representation change to disrupt the original text-form trigger. The sole positive-shift exception is CWE-327 ($\oplus$, $\Delta=+0.103$), which warrants closer examination. In the OWASP Java benchmark, vulnerable CWE-327 samples invoke weak algorithms through factory methods (e.g., \texttt{Cipher.getInstance("DES/\allowbreak CBC/\allowbreak PKCS5Padding")}), while safe samples use strong algorithms (e.g., \texttt{Cipher.getInstance("AES/\allowbreak GCM/\allowbreak NOPADDING")}). The algorithm identifier is a \emph{string-literal argument}, not a class-type token; AST linearisation preserves it unchanged as a \texttt{String} literal node.

Crucially, the aggregate $+0.103$ shift is not uniform across safe samples. Stratifying the 116~negatives by cipher specification (Table~\ref{tab:cwe327-strata}) reveals that the overall positive shift is dominated by a single subgroup:

\begin{table}[htbp]
\centering
\footnotesize
\setlength{\tabcolsep}{3.5pt}
\caption{CWE-327 FPR by cipher-specification stratum (safe samples only, OWASP Java). $\Delta$ = FPR\textsubscript{AST} $-$ FPR\textsubscript{Text} for Text Pilot.}
\label{tab:cwe327-strata}
\resizebox{\columnwidth}{!}{%
\begin{tabular}{lrcccc}
\toprule
\textbf{Stratum} & \textbf{N} & \textbf{FPR\textsubscript{ZS}} & \textbf{FPR\textsubscript{Text}} & \textbf{FPR\textsubscript{AST}} & \textbf{$\Delta$} \\
\midrule
\texttt{AES/GCM}       & 40 & 0.200 & 0.100 & 0.475 & $+$0.375 \\
\texttt{AES/CCM}       & 26 & 0.846 & 0.962 & 0.808 & $-$0.154 \\
Variable-mediated       & 27 & 0.926 & 0.815 & 0.889 & $+$0.074 \\
Provider-indirect       & 23 & 0.304 & 0.826 & 0.783 & $-$0.043 \\
\midrule
\textbf{All CWE-327}   & \textbf{116} & \textbf{0.534} & \textbf{0.603} & \textbf{0.707} & $\mathbf{+}$\textbf{0.103} \\
\bottomrule
\end{tabular}}
\end{table}

The \texttt{AES/GCM} subgroup drives the entire positive shift ($\Delta=+0.375$): the text model attains unusually low FPR (0.100) on these samples, but this drops to~0.475 under AST input---even though the key literal remains available under AST input. Meanwhile, \texttt{AES/CCM} moves in the opposite direction ($\Delta=-0.154$). This decomposition shows that CWE-327's aggregate exception is not a uniform literal-preservation effect, but a mixture of subgroup-specific shifts in which AST appears to suppress some text-specific cues while also disrupting others on which the text model relied.

Two non-exclusive explanations are consistent with these observations: (1)~text fine-tuning learned useful co-occurring context (Java syntax patterns, import structure) that helps identify GCM configurations as safe, and AST strips this context while preserving the literal alone; (2)~the text model's low GCM FPR itself reflects a text-specific shortcut that happened to align with the safe label, and AST disrupts this shortcut along with others. The present data do not distinguish between these accounts. Uneven training-time coverage of different algorithm/mode combinations in NIST Juliet may also contribute to the GCM--CCM divergence. Regardless of the causal mechanism, the decomposition refines the boundary condition: the CWE-327 exception does not falsify the broader surface-cue account, but shows that representation change reshapes the cue landscape non-uniformly, with net direction depending on which text-level patterns dominate in each subgroup, including both harmful shortcuts and patterns that happen to align with the label.

\textbf{Contrast with AST fine-tuning.} The per-CWE heterogeneity observed under the text-trained cross-representation probe does not appear under AST fine-tuning. AST Pilot FPR remains uniformly near ceiling across all 11 categories (0.969--1.000; Table~\ref{tab:finetune-cwe}), with no gradient comparable to the H/M/L/\,$\oplus$ pattern in Table~\ref{tab:cwe-lexical-tiers}. This lack of per-CWE differentiation is consistent with the model relying on cues introduced broadly by flat AST linearisation rather than on CWE-specific lexical triggers.

\textbf{Section synthesis: shortcut anchor layers as an organising hypothesis.}
The preceding analyses---FPR escalation under text fine-tuning, the cross-representation probe's substantial FPR reduction, the asymmetric format-mismatch effect, cross-family replication on Llama, the per-CWE gradient, and the CWE-327 subgroup decomposition---can be unified under a single organising hypothesis. Text fine-tuning appears to rely more heavily on lexical co-occurrence patterns, which are more strongly disrupted by AST serialisation, whereas AST fine-tuning may shift reliance toward cues aligned with structural regularities, some of which remain partially recoverable from raw source text. The CWE-327 subgroup analysis (Table~\ref{tab:cwe327-strata}) illustrates a further nuance: AST's disruption of text-level cues is not uniformly beneficial---different sub-patterns within the same CWE can move in opposite directions, depending on which cues dominate. Taken together, these results suggest that \textbf{AST does not eliminate shortcut dependence}, but changes the mix of cues on which the model relies. Despite these FPR pathologies, AST Pilot still improves aggregate F1 over zero-shot (0.677/0.680 vs.\ 0.653), indicating that task-format adaptation can coexist with severe false-positive behaviour. Whether the anchor-layer pattern generalises beyond Java is examined in the cross-language replication below (\S\ref{sec:python-replication}).

For low-resource CWEs such as CWE-327 and CWE-328, the absence of consistent AST gains suggests that sample scarcity may limit the extent to which representation change alone can help; we return to this limitation in the concluding discussion.

\subsection{Cross-Language Replication: BenchmarkPython}
\label{sec:python-replication}

To test whether the surface-cue memorisation account is specific to Java or reflects a broader property of cross-language deployment, we replicated the key conditions on BenchmarkPython (v0.1; Flask-based Python), using the same Qwen3-8B Text Pilot model trained on C/C++ NIST Juliet. Of the 1,230 raw benchmark samples, 122 (9.9\%) failed Semgrep AST conversion and were excluded; to preserve a fair cross-representation comparison, the resulting 1,108-sample subset (420 vulnerable, 688 benign; 14 CWEs) was used as the canonical evaluation set for \emph{all} Python conditions, including text-based inference. The exclusions are not uniformly distributed across CWEs (per-CWE rates range from 2.6\% to 24.1\%); their full per-CWE and per-label breakdown is reported in the replication package. Accordingly, the Python results are best interpreted as applying to the AST-parseable subset of the benchmark. Python differs substantially from Java in surface realisation, including indentation-based block structure, the absence of type declarations, and distinct library idioms. Accordingly, convergent findings across the two languages would be unlikely to arise solely from Java-specific surface effects.

\begin{table}[htbp]
\centering
\caption{Cross-Language Replication on BenchmarkPython (v0.1, N=1,108, 14 CWEs). Same Qwen3-8B Text Pilot weights as the Java experiments (no retraining). Negative sample count: 688.}
\label{tab:python-replication}
\begin{tabular}{llccc}
\toprule
\textbf{Condition} & \textbf{Input} & \textbf{FPR} & \textbf{Recall} & \textbf{F1} \\
\midrule
Zero-Shot                      & Text & 0.703 & 0.893 & 0.586 \\
Zero-Shot                      & AST  & 0.999 & 1.000 & 0.550 \\
Text Pilot                     & Text & 0.858 & 0.976 & 0.578 \\
\textit{Text Pilot (no retrain)} & AST & \textbf{0.554} & 0.698 & 0.536 \\
\midrule
AST Pilot                      & AST  & 0.986 & 0.986 & 0.548 \\
\textit{AST Pilot (no retrain)} & Text & 0.874 & 0.857 & 0.521 \\
\midrule
\multicolumn{2}{l}{\textit{Reference: Java Text Pilot on AST}} & 0.583 & 0.723 & 0.637 \\
\bottomrule
\end{tabular}
\end{table}

\begin{figure}[htbp]
  \centering
  \includegraphics[width=\linewidth]{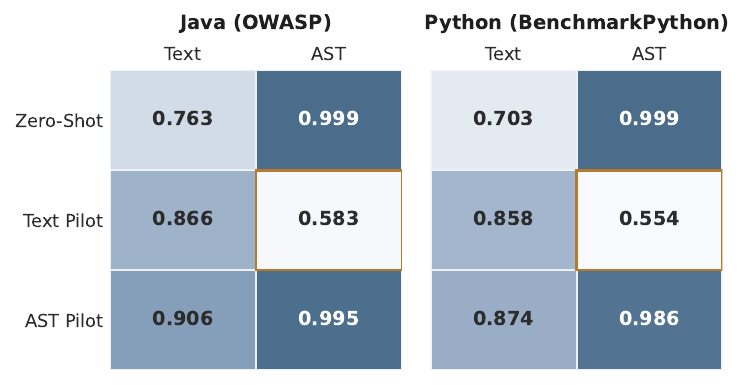}
  \caption{FPR across the training-format $\times$ inference-format matrix (Qwen3-8B, pilot scale), shown separately for Java (left) and Python (right). The dark-bordered cell---Text Pilot on AST input---is the only condition that substantially reduces FPR. All other cells cluster near 0.9--1.0. The two panels exhibit near-identical patterns; the cross-language gap is at most 6.0pp in every cell.}
  \label{fig:fpr-matrix}
\end{figure}

Table~\ref{tab:python-replication} summarises the Python results. The same qualitative progression observed on Java reappears: zero-shot text FPR is already high (0.703), zero-shot AST pushes it to near-ceiling (0.999), and text fine-tuning inflates it further (0.858). The cross-representation probe---applying the same Text Pilot weights to AST-encoded Python samples without retraining---reduces FPR to 0.554, closely matching the Java result (0.583). Figure~\ref{fig:fpr-matrix} places these results alongside the Java data in the full $3 \times 2$ training-format $\times$ inference-format matrix; the cross-language gap in each cell is at most 6.0pp (zero-shot text), indicating convergence across every experimental condition.

\textbf{Headline finding: Python--Java convergence under the AST probe weakens a purely target-language surface-shift account.}
On BenchmarkPython, the cross-representation probe yields FPR$=$0.554, only \textbf{2.9pp below the Java result (0.583)}, despite the substantial surface differences between Python and Java. If target-language surface mismatch were the dominant factor, one would expect a larger separation between the two probe results. Instead, their convergence suggests that once AST encoding disrupts source-form lexical trigger patterns, residual FPR is influenced more by model-side effects of fine-tuning than by target-language surface differences alone. Text Pilot fine-tuning also inflates FPR in both languages to a comparable extent (Python: $+0.155$; Java: $+0.103$), which is consistent with the inflation being driven by shortcut-prone cues inherited from C/C++ fine-tuning rather than by idiosyncratic properties of the evaluation language. This aggregate pattern is especially vivid at the per-CWE level---CWE-330, for instance, exhibits a near-perfect inject--overwrite--strip--restore cycle---but we first complete the global picture before turning to individual cases.

\textbf{AST input produces a more detector-like FPR--Recall trade-off.}
Under text input, Text Pilot achieves Recall$=$0.976 at FPR$=$0.858---near-complete detection at the cost of flagging 85.8\% of safe Python code. Under AST input, Recall falls to 0.698 while FPR drops to 0.554. This 27.8pp Recall reduction, together with the substantial FPR decrease, is consistent with a shift away from near-all-positive collapse toward more discriminative behaviour. The text-input model behaves like a degenerate high-recall, high-FPR classifier; the AST-input model operates in a less collapsed regime, closer to a discriminative detector.

Two per-CWE case studies illustrate how the cross-language surface-cue pattern manifests at fine granularity, bracketing the two extremes of the cue-dependence spectrum.

\textbf{CWE-330: inject--overwrite--strip--restore chain.}
CWE-330 provides the sharpest per-CWE illustration. Zero-shot text FPR$=$0.052 (low baseline FPR); after text fine-tuning, FPR rises to 0.562 (C/C++ triggers \emph{overwrite} the low-FPR baseline); under AST input, FPR \emph{restores} to 0.052, returning to the zero-shot level. This four-step sequence is highly consistent with text fine-tuning inserting shortcut triggers on top of a pre-existing low-FPR baseline rather than replacing it. Full per-CWE FPR data for BenchmarkPython are reported in Appendix Table~\ref{tab:python-cwe-fpr}.

\textbf{CWE-22 (Path Traversal): a format-sensitive boundary case.}
Not all CWEs exhibit the clean pattern of CWE-330. CWE-22 is unusually sensitive to representation switching in \emph{both} directions (Text Pilot$\to$AST: $1.000\to0.591$; AST Pilot$\to$Text: $0.886\to0.045$, Recall$=$0.048). Because safe and vulnerable path-traversal samples share many surface APIs, discriminative signal appears to depend more on how user input, path construction, and validation logic are organised within the representation than on individual lexical tokens. The positive triggers learned under each format are accordingly highly format-dependent and are substantially disrupted by representation switching.

\textbf{Cross-language support for the anchor-layer hypothesis.}
The BenchmarkPython replication exhibits the same directional asymmetry observed on Java: the Text$\to$AST mismatch ($\Delta=-0.30$) is substantially larger than the AST$\to$Text mismatch ($\Delta=-0.112$). This convergence across two syntactically distant target languages lends cross-language support to the organising hypothesis introduced in \S\ref{sec:representation-impact}: the two fine-tuning regimes appear to anchor shortcut-prone cues at different representational layers, with text fine-tuning relying more heavily on source-form lexical co-occurrence patterns that AST serialisation disrupts. The cross-language FPR gap under AST-FT AST conditions (Java~0.995; Python~0.986; gap$=$0.9pp) is tighter than under the Text-FT AST probe (gap$=$2.9pp), which is consistent with AST encoding reducing some language-specific surface variation in how cues are presented---though this should not be overinterpreted as direct evidence that identical structural shortcuts fire across languages.

\section{Conclusion}

This work offers, to our knowledge, the first systematic characterisation of how training-time
and inference-time code representation jointly determine false positive
dynamics in cross-language LLM vulnerability detection. Converging evidence at three levels of
analysis---aggregate model behaviour, per-CWE heterogeneity, and individual
instance counterfactuals---supports \textbf{surface-cue memorisation} as
the primary explanation for the observed false-positive behaviour: text fine-tuning encodes C/C++-specific API names,
string patterns, and syntactic idioms as vulnerability indicators;
inference-time representation determines whether these cues fire on target
input. The cross-representation probe---applying
text-trained weights to AST input without retraining---reduces FPR from
0.866 to 0.583 (Qwen3-8B) and from 0.983 to 0.905
(Llama~3.1-8B-Instruct), while the zero-shot model on AST achieves
FPR$=$0.999, suggesting the reduction is a joint effect of fine-tuning
calibration and AST's disruption of surface-cue triggers. The boundary
condition---AST fine-tuning (Qwen FPR$=$0.995; Llama FPR$=$1.000)---argues against the view that ``AST is just
better'', suggesting instead that flat linearisation introduces structural surface
cues the model similarly memorises.

Notably, in the Text Pilot $\times$ AST probe condition, FPR (0.583) falls
\emph{below} the zero-shot text baseline (0.763), suggesting that text
fine-tuning captures not only representation-specific shortcut cues but also
some signal that remains useful across representations. AST input appears to
disrupt the shortcut-like component more strongly than the task-relevant
component. However, the accompanying Recall reduction
(0.948$\to$0.723) indicates that part of the disrupted surface information
had supported not only false triggering but also correct detections. The CWE-327 subgroup decomposition (Table~\ref{tab:cwe327-strata}) provides per-CWE evidence for the same trade-off on the FPR side: AST disrupts text-specific patterns in a largely non-selective manner, reducing FPR for most subgroups but increasing it in cases where the text model had relied on patterns that, in practice, helped suppress false positives on safe instances. Disentangling these two putative signal
sources---e.g.\ via probing classifiers or causal intervention on
internal representations---is a promising direction for future work.

All four qualitative patterns replicate across both model families, and the
BenchmarkPython replication provides particularly strong evidence against a pure domain-shift
account: the AST probe achieves FPR$=$0.554 on Python---just 2.9pp from Java
(0.583)---despite substantial surface-syntax differences. The CWE-330
inject--overwrite--strip--restore chain
($0.052\!\to\!0.562\!\to\!0.052$) directly illustrates that fine-tuning
overlays shortcut triggers on a pre-existing capacity for lower-FPR judgment.

\textbf{Limitations and future directions.}
First, analysis is restricted to intra-procedural code units;
inter-procedural vulnerabilities spanning multiple functions or files are
out of scope. Second, experiments are limited to 8B-parameter models; whether the same representation-dependent false-positive patterns persist at larger
scales---particularly with respect to cross-domain calibration---remains an open question. Third,
evaluations cover two benchmarks (OWASP v1.2 and BenchmarkPython v0.1) and
two target languages; broader evaluation across additional languages and
real-world corpora is needed. Fourth, all experiments use greedy decoding
and a fixed strict-JSON system prompt; different decoding strategies or
prompt templates may shift absolute FPR values. While we expect the main
directional patterns to be less sensitive than the absolute rates, this
has not been directly tested here. Fifth, the cross-language replication uses Qwen3-8B only; Llama~3.1-8B-Instruct was evaluated on Java but not on BenchmarkPython due to compute constraints, so the cross-family evidence is limited to Java. Sixth, the full-scale text and AST training sets differ slightly in composition (569 samples, 0.8\%, failed AST conversion; see \S\ref{sec:dataset-prep}), and the BenchmarkPython evaluation is defined on the AST-parseable subset (1,108 of 1,230 samples), with non-uniform per-CWE exclusion rates. Because the core cross-representation analyses rest on the pilot split---where text and AST training sets differ by only one sample---we do not expect these composition differences to materially affect the main conclusions, but they represent a boundary on strict comparability at full scale. Seventh, the cross-representation probe
requires access to model weights for adapter loading, precluding
evaluation on closed-source models (e.g., GPT-4); whether large proprietary
models exhibit the same cross-representation behaviour remains
unknown. On a related note, simple token-level
metrics---cross-language lexical overlap (Jaccard) and AST token survival
rate---do not predict per-CWE $\Delta$FPR in the pooled Java+Python analysis
($R^2 = 0.013$, $p = 0.584$ and $R^2 = 0.041$, $p = 0.33$, respectively),
suggesting that the relevant surface-cue effects operate at a broader pattern
level rather than being captured by identifier-level overlap alone;
predicting which CWEs are most susceptible will likely require
model-internal analysis such as probing classifiers or attention
attribution. Moving forward, we recommend exploring richer structural
representations such as Program Dependence Graphs (PDGs) and Code Property
Graphs (CPGs), which encode semantic data-flow and control-flow
relationships that raw AST cannot capture. Additionally, training regimes
that explicitly supervise on target-domain negative samples are a promising
direction for reducing false positives to practically usable levels.

\textbf{Practical implications.} For practitioners, our findings carry three implications: cross-language SAST evaluation should prioritise false-positive control rather than relying on aggregate metrics such as F1 alone; representation disagreement between text and AST paths can serve as a practical triage signal; and richer semantic representations are a more promising direction than na\"ive AST fine-tuning. Concretely, the cross-representation probe can serve as a \emph{pre-deployment consistency gate}: before releasing a fine-tuned detector, run every candidate alert through both the text and AST input paths; alerts flagged under text but cleared under AST are enriched for surface-cue-driven false positives and may be suppressed or routed to human review. In our experiments, this ``representation-agreement'' filter would have eliminated 427 of 1{,}147 false positives (37.2\%) with no retraining, at the cost of a recall reduction from 0.948 to 0.723---a trade-off suited to false-positive-sensitive deployment scenarios such as security audits and high-cost human triage, where each unnecessary alert carries substantial review overhead. The effect is directionally consistent across both model families (Llama~3.1-8B-Instruct: $\Delta=-0.078$) and consistent across the two evaluated language settings (BenchmarkPython $\Delta=-0.304$ vs.\ Java $\Delta=-0.282$). The gate could be explored as a post-hoc alert-filtering layer in CI/CD-based SAST pipelines: an alert is promoted only when both representations agree on ``vulnerable,'' while disagreement triggers a lower-confidence flag. This conservative AND-gate strategy prioritises low FPR over maximal recall, and is therefore best suited to deployment settings where false-positive cost substantially exceeds miss cost.


\appendix
\section{Detailed Baseline Performance Metrics}
\label{sec:appendix-a}

This section provides the unabridged performance results of the baseline models on the OWASP Benchmark, disaggregated by each Common Weakness Enumeration (CWE) category.

\subsection{Detailed Performance with Confusion Matrices}

Table~\ref{tab:detailed-cwe-performance} provides the full confusion matrix breakdown for Qwen3-8B zero-shot baseline, enabling analysis of true/false positive patterns across vulnerability categories.

\begin{table*}[htbp]
\centering
\caption{Detailed Per-CWE Performance of Qwen3-8B Zero-Shot Baseline with Confusion Matrices (TP/TN/FP/FN)}
\label{tab:detailed-cwe-performance}
\footnotesize
\setlength{\tabcolsep}{4pt}
\begin{tabular}{llr|rrrr|rrr}
\toprule
\textbf{CWE} & \textbf{Category} & \textbf{N} & \textbf{TP} & \textbf{TN} & \textbf{FP} & \textbf{FN} & \textbf{Rec} & \textbf{Prec} & \textbf{F1} \\
\midrule
89  & sqli       & 504 & 270 & 2   & 230 & 2   & 0.9926 & 0.5400 & 0.6995 \\
22  & pathtraver & 268 & 131 & 4   & 131 & 2   & 0.9850 & 0.5000 & 0.6633 \\
78  & cmdi       & 251 & 126 & 0   & 125 & 0   & 1.0000 & 0.5020 & 0.6684 \\
90  & ldapi      & 59  & 27  & 1   & 31  & 0   & 1.0000 & 0.4655 & 0.6353 \\
643 & xpathi     & 35  & 15  & 0   & 20  & 0   & 1.0000 & 0.4286 & 0.6000 \\
\midrule
328 & hash       & 236 & 116 & 33  & 74  & 13  & 0.8992 & 0.6105 & 0.7273 \\
327 & crypto     & 246 & 116 & 54  & 62  & 14  & 0.8923 & 0.6517 & 0.7532 \\
\midrule
614 & securecookie& 67  & 3   & 31  & 0   & 33  & 0.0833 & 1.0000 & 0.1538 \\
79  & xss        & 455 & 217 & 49  & 160 & 29  & 0.8821 & 0.5756 & 0.6966 \\
330 & weakrand   & 493 & 133 & 108 & 167 & 85  & 0.6101 & 0.4433 & 0.5135 \\
501 & trustbound & 126 & 21  & 32  & 11  & 62  & 0.2530 & 0.6562 & 0.3652 \\
\midrule
\textbf{All} & & \textbf{2740} & \textbf{1175} & \textbf{314} & \textbf{1011} & \textbf{240} & \textbf{0.8304} & \textbf{0.5375} & \textbf{0.6526} \\
\bottomrule
\end{tabular}
\end{table*}

\subsection{Per-CWE FPR: Cross-Representation Probe Breakdown}
\label{appendix:cwe-fpr}

Table~\ref{tab:cwe-fpr-crossrepr} reports the false positive rate for each CWE category across key experimental conditions for Qwen3-8B (zero-shot, Text Pilot, cross-representation probe, AST fine-tuned) and Llama~3.1-8B-Instruct (Text Pilot and cross-representation probe).
The $\Delta$ columns report $\text{FPR}_\text{AST\,input} - \text{FPR}_\text{Text\,input}$ for the respective Text Pilot model; negative values indicate a reduction in FPR when switching to AST input.

\begin{table*}[htbp]
\centering
\caption{Per-CWE FPR across training and evaluation conditions on OWASP Benchmark v1.2 (Java).
    The native eval block shows each model evaluated on its training format;
    AST Full uses the full-dataset AST-trained model.
    The cross-repr probe blocks show the effect of switching evaluation input
    from text to AST without retraining.
    $\Delta$ = FPR(Text Pilot, AST) $-$ FPR(Text Pilot, Text); negative values indicate
    AST input reduces FPR.}
\label{tab:cwe-fpr-crossrepr}
\footnotesize
\setlength{\tabcolsep}{3.5pt}
\begin{tabular}{llccc|cc|ccc}
\toprule
& & \multicolumn{3}{c|}{\textbf{Qwen3-8B (native eval)}} & \multicolumn{2}{c|}{\textbf{Qwen3-8B cross-repr}} & \multicolumn{3}{c}{\textbf{Llama~3.1-8B cross-repr}} \\
& & \textbf{Zero-Shot} & \textbf{Text Pilot} & \textbf{AST Full} & \textbf{Text Pilot} & & \textbf{Text Pilot} & \textbf{Text Pilot} & \\
\textbf{CWE} & \textbf{Category} & \textbf{Text} & \textbf{Text} & \textbf{AST} & \textbf{AST} & \textbf{$\Delta$} & \textbf{Text} & \textbf{AST} & \textbf{$\Delta$} \\
\midrule
CWE-22  & Path traversal   & 0.970 & 0.933 & 0.978 & 0.526 & $-$0.407 & 1.000 & 1.000 & $\phantom{+}$0.000 \\
CWE-78  & Command inject.  & 1.000 & 0.992 & 1.000 & 0.888 & $-$0.104 & 1.000 & 1.000 & $\phantom{+}$0.000 \\
CWE-79  & XSS              & 0.766 & 0.909 & 0.967 & 0.531 & $-$0.378 & 1.000 & 1.000 & $\phantom{+}$0.000 \\
CWE-89  & SQL injection    & 0.991 & 0.935 & 0.927 & 0.797 & $-$0.138 & 0.996 & 0.966 & $-$0.030 \\
CWE-90  & LDAP injection   & 0.969 & 0.750 & 0.938 & 0.688 & $-$0.062 & 1.000 & 1.000 & $\phantom{+}$0.000 \\
CWE-327 & Weak crypto alg. & 0.534 & 0.603 & 0.983 & 0.707 & $+$0.103 & 1.000 & 0.638 & $-$0.362 \\
CWE-328 & Weak hash        & 0.692 & 0.944 & 0.972 & 0.682 & $-$0.262 & 1.000 & 0.738 & $-$0.262 \\
CWE-330 & Insuff.\ random. & 0.607 & 0.771 & 0.993 & 0.244 & $-$0.527 & 0.989 & 0.909 & $-$0.080 \\
CWE-501 & Trust boundary   & 0.256 & 0.884 & 0.953 & 0.651 & $-$0.233 & 1.000 & 1.000 & $\phantom{+}$0.000 \\
CWE-614 & Sensitive cookie & 0.000 & 0.806 & 0.968 & 0.387 & $-$0.419 & 0.387 & 0.258 & $-$0.129 \\
CWE-643 & XPath injection  & 1.000 & 1.000 & 0.950 & 0.550 & $-$0.450 & 1.000 & 1.000 & $\phantom{+}$0.000 \\
\midrule
\textbf{Overall} & & 0.763 & 0.866 & 0.970 & 0.583 & $-$0.282 & 0.983 & 0.905 & $-$0.078 \\
\bottomrule
\end{tabular}
\end{table*}

\subsection{Per-CWE FPR: BenchmarkPython Cross-Representation Probe}
\label{appendix:python-cwe-fpr}

Table~\ref{tab:python-cwe-fpr} reports the per-CWE false positive rate on BenchmarkPython (v0.1, 14~CWEs) across all training and evaluation conditions: Zero-Shot, Text Pilot, and AST Pilot under native evaluation, plus both cross-representation probes (Text Pilot on AST input and AST Pilot on Text input). The $\Delta$ column reports FPR(Text Pilot, AST) $-$ FPR(Text Pilot, Text). This table provides the per-CWE evidence underlying the CWE-330 inject--overwrite--strip--restore chain discussed in Section~\ref{sec:python-replication}.

\begin{table*}[htbp]
\centering
\caption{Per-CWE FPR on BenchmarkPython (Qwen3-8B). The native eval block shows each model tested on its training format. The cross-repr probe blocks show evaluation input switched without retraining. For each probe, $\Delta$ = FPR(cross-repr) $-$ FPR(native); negative values indicate the format switch reduces FPR. Sorted by Text Pilot $\Delta$. Three Python-specific CWEs (CWE-502, CWE-601, CWE-94) have no Java counterpart in OWASP.}
\label{tab:python-cwe-fpr}
\footnotesize
\setlength{\tabcolsep}{2.5pt}
\begin{tabular}{llcccc|cc|cc}
\toprule
& & & \multicolumn{3}{c|}{\textbf{Native eval}} & \multicolumn{2}{c|}{\textbf{Text Pilot cross-repr}} & \multicolumn{2}{c}{\textbf{AST Pilot cross-repr}} \\
& & & \textbf{Zero-Shot} & \textbf{Text Pilot} & \textbf{AST Pilot} & \textbf{Text Pilot} & & \textbf{AST Pilot} & \\
\textbf{CWE} & \textbf{Category} & \textbf{N} & \textbf{Text} & \textbf{Text} & \textbf{AST} & \textbf{AST} & \textbf{$\Delta$} & \textbf{Text} & \textbf{$\Delta$} \\
\midrule
CWE-501 & Trust boundary   &  31 & 0.857 & 1.000 & 1.000 & 0.286 & $-$0.714 & 1.000 & $\phantom{+}$0.000 \\
CWE-330 & Insuff.\ random. & 302 & 0.052 & 0.562 & 1.000 & 0.052 & $-$0.509 & 0.995 & $-$0.005 \\
CWE-22  & Path traversal   & 150 & 1.000 & 1.000 & 0.886 & 0.591 & $-$0.409 & 0.045 & $-$0.841 \\
CWE-90  & LDAP injection   &  22 & 1.000 & 0.900 & 1.000 & 0.500 & $-$0.400 & 1.000 & $\phantom{+}$0.000 \\
CWE-502 & Deserialization  &  49 & 1.000 & 1.000 & 1.000 & 0.719 & $-$0.281 & 1.000 & $\phantom{+}$0.000 \\
CWE-601 & Open redirect    &  29 & 1.000 & 0.938 & 1.000 & 0.750 & $-$0.188 & 1.000 & $\phantom{+}$0.000 \\
CWE-643 & XPath injection  & 161 & 1.000 & 0.983 & 1.000 & 0.809 & $-$0.174 & 1.000 & $\phantom{+}$0.000 \\
CWE-328 & Weak hash        & 143 & 1.000 & 0.974 & 1.000 & 0.803 & $-$0.171 & 1.000 & $\phantom{+}$0.000 \\
CWE-78  & Command inject.  &  18 & 1.000 & 1.000 & 1.000 & 0.857 & $-$0.143 & 1.000 & $\phantom{+}$0.000 \\
CWE-89  & SQL injection    &  14 & 1.000 & 1.000 & 1.000 & 0.889 & $-$0.111 & 1.000 & $\phantom{+}$0.000 \\
CWE-94  & Code injection   &  43 & 1.000 & 1.000 & 1.000 & 0.920 & $-$0.080 & 1.000 & $\phantom{+}$0.000 \\
CWE-79  & XSS              &  84 & 0.963 & 1.000 & 1.000 & 0.944 & $-$0.056 & 0.963 & $-$0.037 \\
CWE-611 & XXE              &  24 & 1.000 & 1.000 & 1.000 & 1.000 & $\phantom{+}$0.000 & 1.000 & $\phantom{+}$0.000 \\
CWE-614 & Sensitive cookie &  38 & 0.933 & 1.000 & 1.000 & 1.000 & $\phantom{+}$0.000 & 1.000 & $\phantom{+}$0.000 \\
\midrule
\textbf{Overall} & & 1108 & 0.703 & 0.858 & 0.986 & 0.554 & $-$0.304 & 0.874 & $-$0.112 \\
\bottomrule
\end{tabular}
\end{table*}

\section{Training Hyperparameters}
\label{appendix:hyperparams}

Table~\ref{tab:hyperparams} summarises the per-run hyperparameter settings for all fine-tuning conditions.

All fine-tuning runs share the same core configuration: AdamW optimiser with cosine schedule (warmup ratio 0.02), no weight decay, and gradient norm clipping at 1.0. The learning rate is $2 \times 10^{-4}$ for both Qwen3-8B and Llama~3.1-8B. LoRA adapters (rank~16, $\alpha$~32, dropout 0.05) target the four attention projections (\texttt{q\_proj}, \texttt{k\_proj}, \texttt{v\_proj}, \texttt{o\_proj}). Training uses BF16 precision with gradient checkpointing; early stopping monitors validation F1 (patience~2). All seeds are fixed at 42.

\begin{table}[htbp]
\centering
\caption{Per-run hyperparameters that differ across training configurations. ``Eff.\ BS'' = per-device batch size $\times$ gradient accumulation steps.}
\label{tab:hyperparams}
\footnotesize
\resizebox{\columnwidth}{!}{%
\begin{tabular}{llcccc}
\toprule
\textbf{Training} & \textbf{Model} & \textbf{Eff.\ BS} & \textbf{Max Len} & \textbf{Epochs} \\
\midrule
Text Pilot  & Qwen3-8B      & $1 \times 4 = 4$  & 2048 & 3 \\
Text Full   & Qwen3-8B      & $4 \times 4 = 16$ & 1024 & 3\textsuperscript{\dag} \\
Text Pilot  & Llama-3.1-8B  & $2 \times 4 = 8$  & 2048 & 3 \\
\midrule
AST Pilot   & Qwen3-8B      & $2 \times 4 = 8$  & 4096 & 3 \\
AST Full    & Qwen3-8B      & $4 \times 4 = 16$ & 4096 & 3\textsuperscript{\dag} \\
AST Pilot   & Llama-3.1-8B  & $2 \times 4 = 8$  & 4096 & 3 \\
\bottomrule
\end{tabular}}
\par\smallskip
\raggedright\scriptsize\textsuperscript{\dag}Configured for 3 epochs; training was terminated manually once loss had largely plateaued. Results are reported from the epoch-1 and epoch-2 checkpoints.
\textsuperscript{\ddag}AST runs use 4096 tokens because S-expression linearisation approximately doubles sequence length relative to raw text.
\end{table}

\section*{Declarations}

\noindent\textbf{Competing interests.}
The authors declare no competing interests.

\noindent\textbf{Data and code availability.}
The original benchmark datasets used in this study (NIST Juliet Test Suite, OWASP Benchmark v1.2, and BenchmarkPython v0.1) are publicly available from their respective sources. A complete replication package, including model predictions, processed evaluation splits, fine-tuned LoRA adapters, training, inference, verification, and analysis scripts, is available from the corresponding author upon reasonable request and will be released publicly upon acceptance.

\bibliographystyle{ACM-Reference-Format}
\bibliography{repcue}


\begin{thebibliography}{20}


\ifx \showCODEN    \undefined \def \showCODEN     #1{\unskip}     \fi
\ifx \showISBNx    \undefined \def \showISBNx     #1{\unskip}     \fi
\ifx \showISBNxiii \undefined \def \showISBNxiii  #1{\unskip}     \fi
\ifx \showISSN     \undefined \def \showISSN      #1{\unskip}     \fi
\ifx \showLCCN     \undefined \def \showLCCN      #1{\unskip}     \fi
\ifx \shownote     \undefined \def \shownote      #1{#1}          \fi
\ifx \showarticletitle \undefined \def \showarticletitle #1{#1}   \fi
\ifx \showURL      \undefined \def \showURL       {\relax}        \fi
\providecommand\bibfield[2]{#2}
\providecommand\bibinfo[2]{#2}
\providecommand\natexlab[1]{#1}
\providecommand\showeprint[2][]{arXiv:#2}

\bibitem[Boland and Black(2012)]%
        {bolandjuliet2012}
\bibfield{author}{\bibinfo{person}{Tim Boland} {and} \bibinfo{person}{Paul~E.
  Black}.} \bibinfo{year}{2012}\natexlab{}.
\newblock \showarticletitle{The {{Juliet}} 1.1 {{C}}/{{C}}++ and {{Java Test
  Suite}}}.
\newblock \bibinfo{journal}{\emph{Computer}} \bibinfo{volume}{45},
  \bibinfo{number}{10} (\bibinfo{year}{2012}), \bibinfo{pages}{83--90}.
\newblock
\href{https://doi.org/10.1109/MC.2012.345}{doi:\nolinkurl{10.1109/MC.2012.345}}


\bibitem[Chakraborty et~al\mbox{.}(2022)]%
        {chakrabortydeep2022}
\bibfield{author}{\bibinfo{person}{Saikat Chakraborty}, \bibinfo{person}{Rahul
  Krishna}, \bibinfo{person}{Yangruibo Ding}, {and} \bibinfo{person}{Baishakhi
  Ray}.} \bibinfo{year}{2022}\natexlab{}.
\newblock \showarticletitle{Deep {{Learning Based Vulnerability Detection}}:
  {{Are We There Yet}}?}
\newblock \bibinfo{journal}{\emph{IEEE Transactions on Software Engineering}}
  \bibinfo{volume}{48}, \bibinfo{number}{9} (\bibinfo{year}{2022}),
  \bibinfo{pages}{3280--3296}.
\newblock
\href{https://doi.org/10.1109/TSE.2021.3087402}{doi:\nolinkurl{10.1109/TSE.2021.3087402}}


\bibitem[Chen et~al\mbox{.}(2023)]%
        {chendiversevul2023}
\bibfield{author}{\bibinfo{person}{Yizheng Chen}, \bibinfo{person}{Zhoujie
  Ding}, \bibinfo{person}{Lamya Alowain}, \bibinfo{person}{Xinyun Chen}, {and}
  \bibinfo{person}{David Wagner}.} \bibinfo{year}{2023}\natexlab{}.
\newblock \showarticletitle{{{DiverseVul}}: {{A New Vulnerable Source Code
  Dataset}} for {{Deep Learning Based Vulnerability Detection}}}. In
  \bibinfo{booktitle}{\emph{Proceedings of the 26th {{International Symposium}}
  on {{Research}} in {{Attacks}}, {{Intrusions}} and {{Defenses}}}}.
  \bibinfo{pages}{654--668}.
\newblock
\href{https://doi.org/10.1145/3607199.3607242}{doi:\nolinkurl{10.1145/3607199.3607242}}


\bibitem[Du et~al\mbox{.}(2023)]%
        {dujoint2023}
\bibfield{author}{\bibinfo{person}{Qianjin Du}, \bibinfo{person}{Shiji Zhou},
  \bibinfo{person}{Xiaohui Kuang}, \bibinfo{person}{Gang Zhao}, {and}
  \bibinfo{person}{Jidong Zhai}.} \bibinfo{year}{2023}\natexlab{}.
\newblock \showarticletitle{Joint {{Geometrical}} and {{Statistical Domain
  Adaptation}} for {{Cross-domain Code Vulnerability Detection}}}. In
  \bibinfo{booktitle}{\emph{Proceedings of the 2023 {{Conference}} on
  {{Empirical Methods}} in {{Natural Language Processing}}}}.
  \bibinfo{pages}{12791--12800}.
\newblock
\href{https://doi.org/10.18653/v1/2023.emnlp-main.788}{doi:\nolinkurl{10.18653/v1/2023.emnlp-main.788}}


\bibitem[Feng et~al\mbox{.}(2020)]%
        {fengcodebert2020}
\bibfield{author}{\bibinfo{person}{Zhangyin Feng}, \bibinfo{person}{Daya Guo},
  \bibinfo{person}{Duyu Tang}, \bibinfo{person}{Nan Duan},
  \bibinfo{person}{Xiaochuan Feng}, \bibinfo{person}{Ming Gong},
  \bibinfo{person}{Linjun Shou}, \bibinfo{person}{Bing Qin},
  \bibinfo{person}{Ting Liu}, \bibinfo{person}{Daxin Jiang}, {and}
  \bibinfo{person}{Ming Zhou}.} \bibinfo{year}{2020}\natexlab{}.
\newblock \showarticletitle{{{CodeBERT}}: {{A Pre-Trained Model}} for
  {{Programming}} and {{Natural Language}}}. In
  \bibinfo{booktitle}{\emph{Findings of the {{Association}} for {{Computational
  Linguistics}}: {{EMNLP}} 2020}}. \bibinfo{pages}{1536--1547}.
\newblock
\href{https://doi.org/10.18653/v1/2020.findings-emnlp.139}{doi:\nolinkurl{10.18653/v1/2020.findings-emnlp.139}}


\bibitem[Fu and Tantithamthavorn(2022)]%
        {fulinevul2022}
\bibfield{author}{\bibinfo{person}{Michael Fu} {and} \bibinfo{person}{Chakkrit
  Tantithamthavorn}.} \bibinfo{year}{2022}\natexlab{}.
\newblock \showarticletitle{{{LineVul}}: {{A Transformer-based Line-Level
  Vulnerability Prediction}}}. In \bibinfo{booktitle}{\emph{Proceedings of the
  19th {{International Conference}} on {{Mining Software Repositories}}}}.
  \bibinfo{pages}{608--620}.
\newblock
\href{https://doi.org/10.1145/3524842.3528452}{doi:\nolinkurl{10.1145/3524842.3528452}}


\bibitem[Gao et~al\mbox{.}(2024)]%
        {gaohow2024}
\bibfield{author}{\bibinfo{person}{Zeyu Gao}, \bibinfo{person}{Hao Wang},
  \bibinfo{person}{Yuchen Zhou}, \bibinfo{person}{Yuandong Ni}, {and}
  \bibinfo{person}{Chao Zhang}.} \bibinfo{year}{2024}\natexlab{}.
\newblock \showarticletitle{How {{Far Have We Gone}} in {{Vulnerability
  Detection Using Large Language Models}}}.
\newblock \bibinfo{journal}{\emph{arXiv preprint arXiv:2311.12420}}
  (\bibinfo{year}{2024}).
\newblock
\showeprint[arxiv]{2311.12420}
\href{https://doi.org/10.48550/arXiv.2311.12420}{doi:\nolinkurl{10.48550/arXiv.2311.12420}}


\bibitem[Geirhos et~al\mbox{.}(2020)]%
        {geirhosshortcut2020}
\bibfield{author}{\bibinfo{person}{Robert Geirhos},
  \bibinfo{person}{J{\"o}rn-Henrik Jacobsen}, \bibinfo{person}{Claudio
  Michaelis}, \bibinfo{person}{Richard Zemel}, \bibinfo{person}{Matthias
  Bethge}, \bibinfo{person}{Bernhard Sch{\"o}lkopf}, {and}
  \bibinfo{person}{Felix~A Wichmann}.} \bibinfo{year}{2020}\natexlab{}.
\newblock \showarticletitle{Shortcut {{Learning}} in {{Deep Neural Networks}}}.
\newblock \bibinfo{journal}{\emph{Nature Machine Intelligence}}
  \bibinfo{volume}{2}, \bibinfo{number}{11} (\bibinfo{year}{2020}),
  \bibinfo{pages}{665--673}.
\newblock
\href{https://doi.org/10.1038/s42256-020-00257-z}{doi:\nolinkurl{10.1038/s42256-020-00257-z}}


\bibitem[Guo et~al\mbox{.}(2022)]%
        {guounixcoder2022}
\bibfield{author}{\bibinfo{person}{Daya Guo}, \bibinfo{person}{Shuai Lu},
  \bibinfo{person}{Nan Duan}, \bibinfo{person}{Yanlin Wang},
  \bibinfo{person}{Ming Zhou}, {and} \bibinfo{person}{Jian Yin}.}
  \bibinfo{year}{2022}\natexlab{}.
\newblock \showarticletitle{{{UniXcoder}}: {{Unified Cross-Modal Pre-training}}
  for {{Code Representation}}}. In \bibinfo{booktitle}{\emph{Proceedings of the
  60th {{Annual Meeting}} of the {{Association}} for {{Computational
  Linguistics}}}}. \bibinfo{pages}{7212--7225}.
\newblock
\href{https://doi.org/10.18653/v1/2022.acl-long.499}{doi:\nolinkurl{10.18653/v1/2022.acl-long.499}}


\bibitem[Harer et~al\mbox{.}(2018)]%
        {harerautomated2018}
\bibfield{author}{\bibinfo{person}{Jacob~A. Harer}, \bibinfo{person}{Louis~Y.
  Kim}, \bibinfo{person}{Rebecca~L. Russell}, \bibinfo{person}{Onur Ozdemir},
  \bibinfo{person}{Leonard~R. Kosta}, \bibinfo{person}{Akshay Rangamani},
  \bibinfo{person}{Lei~H. Hamilton}, \bibinfo{person}{Gabriel~I. Centeno},
  \bibinfo{person}{Jonathan~R. Key}, \bibinfo{person}{Paul~M. Ellingwood},
  \bibinfo{person}{Marc~W. McConley}, \bibinfo{person}{Jeffrey~M. Mattson},
  \bibinfo{person}{Andy Long}, \bibinfo{person}{Chris Enck},
  \bibinfo{person}{Steven Miley}, {and} \bibinfo{person}{Tomo Lazovich}.}
  \bibinfo{year}{2018}\natexlab{}.
\newblock \showarticletitle{Automated {{Software Vulnerability Detection}} with
  {{Machine Learning}}}.
\newblock \bibinfo{journal}{\emph{arXiv preprint arXiv:1803.04497}}
  (\bibinfo{year}{2018}).
\newblock
\showeprint[arxiv]{1803.04497}
\href{https://doi.org/10.48550/arXiv.1803.04497}{doi:\nolinkurl{10.48550/arXiv.1803.04497}}


\bibitem[Jiang et~al\mbox{.}(2024)]%
        {jianginvestigating2024}
\bibfield{author}{\bibinfo{person}{Xuefeng Jiang}, \bibinfo{person}{Lvhua Wu},
  \bibinfo{person}{Sheng Sun}, \bibinfo{person}{Jia Li},
  \bibinfo{person}{Jingjing Xue}, \bibinfo{person}{Yuwei Wang},
  \bibinfo{person}{Tingting Wu}, {and} \bibinfo{person}{Min Liu}.}
  \bibinfo{year}{2024}\natexlab{}.
\newblock \bibinfo{title}{Investigating {{Large Language Models}} for {{Code
  Vulnerability Detection}}: {{An Experimental Study}}}.
\newblock
\href{https://doi.org/10.48550/arXiv.2412.18260}{doi:\nolinkurl{10.48550/arXiv.2412.18260}}


\bibitem[Li et~al\mbox{.}(2024)]%
        {liunveiling2024}
\bibfield{author}{\bibinfo{person}{Zhiming Li}, \bibinfo{person}{Yanzhou Li},
  \bibinfo{person}{Tianlin Li}, \bibinfo{person}{Mengnan Du},
  \bibinfo{person}{Bozhi Wu}, \bibinfo{person}{Yushi Cao},
  \bibinfo{person}{Junzhe Jiang}, {and} \bibinfo{person}{Yang Liu}.}
  \bibinfo{year}{2024}\natexlab{}.
\newblock \showarticletitle{Unveiling {{Project-Specific Bias}} in {{Neural
  Code Models}}}. In \bibinfo{booktitle}{\emph{Proceedings of the 2024 {{Joint
  International Conference}} on {{Computational Linguistics}}, {{Language
  Resources}} and {{Evaluation}} ({{LREC-COLING}} 2024)}}.
  \bibinfo{pages}{17205--17216}.
\newblock
\href{https://doi.org/10.18653/v1/2024.lrec-main.1494}{doi:\nolinkurl{10.18653/v1/2024.lrec-main.1494}}


\bibitem[Li et~al\mbox{.}(2022)]%
        {lisysevr2022}
\bibfield{author}{\bibinfo{person}{Zhen Li}, \bibinfo{person}{Deqing Zou},
  \bibinfo{person}{Shouhuai Xu}, \bibinfo{person}{Hai Jin},
  \bibinfo{person}{Hanchao Qi}, {and} \bibinfo{person}{Jie Hu}.}
  \bibinfo{year}{2022}\natexlab{}.
\newblock \showarticletitle{{{SySeVR}}: {{A Framework}} for {{Using Deep
  Learning}} to {{Detect Software Vulnerabilities}}}.
\newblock \bibinfo{journal}{\emph{IEEE Transactions on Dependable and Secure
  Computing}} \bibinfo{volume}{19}, \bibinfo{number}{4} (\bibinfo{year}{2022}),
  \bibinfo{pages}{2244--2258}.
\newblock
\href{https://doi.org/10.1109/TDSC.2021.3051525}{doi:\nolinkurl{10.1109/TDSC.2021.3051525}}


\bibitem[Liu et~al\mbox{.}(2024)]%
        {liuvuldetectbench2024}
\bibfield{author}{\bibinfo{person}{Yu Liu}, \bibinfo{person}{Lang Gao},
  \bibinfo{person}{Mingxin Yang}, \bibinfo{person}{Yu Xie},
  \bibinfo{person}{Ping Chen}, \bibinfo{person}{Xiaojin Zhang}, {and}
  \bibinfo{person}{Wei Chen}.} \bibinfo{year}{2024}\natexlab{}.
\newblock \showarticletitle{{{VulDetectBench}}: {{Evaluating}} the {{Deep
  Capability}} of {{Vulnerability Detection}} with {{Large Language Models}}}.
\newblock \bibinfo{journal}{\emph{arXiv preprint arXiv:2406.07595}}
  (\bibinfo{year}{2024}).
\newblock
\showeprint[arxiv]{2406.07595}
\href{https://doi.org/10.48550/arXiv.2406.07595}{doi:\nolinkurl{10.48550/arXiv.2406.07595}}


\bibitem[Mirsky et~al\mbox{.}(2023)]%
        {mirskyvulchecker2023}
\bibfield{author}{\bibinfo{person}{Yisroel Mirsky}, \bibinfo{person}{George
  Macon}, \bibinfo{person}{Michael Brown}, \bibinfo{person}{Carter Yagemann},
  \bibinfo{person}{Matthew Pruett}, \bibinfo{person}{Evan Downing},
  \bibinfo{person}{Sukarno Mertoguno}, {and} \bibinfo{person}{Wenke Lee}.}
  \bibinfo{year}{2023}\natexlab{}.
\newblock \showarticletitle{{{VulChecker}}: {{Graph-based Vulnerability
  Localization}} in {{Source Code}}}. In \bibinfo{booktitle}{\emph{32nd
  {{USENIX Security Symposium}}}}. \bibinfo{pages}{2041--2058}.
\newblock
\href{https://doi.org/10.5555/3620237.3620604}{doi:\nolinkurl{10.5555/3620237.3620604}}


\bibitem[Nguyen et~al\mbox{.}(2025)]%
        {nguyenmulvuln2025}
\bibfield{author}{\bibinfo{person}{Van Nguyen}, \bibinfo{person}{Surya Nepal},
  \bibinfo{person}{Xingliang Yuan}, \bibinfo{person}{Tingmin Wu},
  \bibinfo{person}{Fengchao Chen}, {and} \bibinfo{person}{Carsten Rudolph}.}
  \bibinfo{year}{2025}\natexlab{}.
\newblock \bibinfo{title}{{{MulVuln}}: {{Enhancing Pre-trained LMs}} with
  {{Shared}} and {{Language-Specific Knowledge}} for {{Multilingual
  Vulnerability Detection}}}.
\newblock
\href{https://doi.org/10.48550/arXiv.2510.04397}{doi:\nolinkurl{10.48550/arXiv.2510.04397}}


\bibitem[Nikitopoulos et~al\mbox{.}(2021)]%
        {nikitopouloscrossvul2021}
\bibfield{author}{\bibinfo{person}{Georgios Nikitopoulos},
  \bibinfo{person}{Konstantina Dritsa}, \bibinfo{person}{Panos Louridas}, {and}
  \bibinfo{person}{Dimitris Mitropoulos}.} \bibinfo{year}{2021}\natexlab{}.
\newblock \showarticletitle{{{CrossVul}}: {{A Cross-Language Vulnerability
  Dataset}} with {{Commit Data}}}. In \bibinfo{booktitle}{\emph{Proceedings of
  the 29th {{ACM Joint Meeting}} on {{European Software Engineering
  Conference}} and {{Symposium}} on the {{Foundations}} of {{Software
  Engineering}}}}. \bibinfo{pages}{1565--1569}.
\newblock
\href{https://doi.org/10.1145/3468264.3473122}{doi:\nolinkurl{10.1145/3468264.3473122}}


\bibitem[Wichers and Maher(2015)]%
        {wichersowasp2015}
\bibfield{author}{\bibinfo{person}{Dave Wichers} {and} \bibinfo{person}{Dave
  Maher}.} \bibinfo{year}{2015}\natexlab{}.
\newblock \bibinfo{title}{{{OWASP Benchmark Project}}}.
\newblock
\urldef\tempurl%
\url{https://owasp.org/www-project-benchmark/}
\showURL{%
\tempurl}


\bibitem[Zhou et~al\mbox{.}(2019)]%
        {zhoudevign2019}
\bibfield{author}{\bibinfo{person}{Yaqin Zhou}, \bibinfo{person}{Shangqing
  Liu}, \bibinfo{person}{Jingkai Siow}, \bibinfo{person}{Xiaoning Du}, {and}
  \bibinfo{person}{Yang Liu}.} \bibinfo{year}{2019}\natexlab{}.
\newblock \showarticletitle{Devign: {{Effective Vulnerability Identification}}
  by {{Learning Comprehensive Program Semantics}} via {{Graph Neural
  Networks}}}. In \bibinfo{booktitle}{\emph{Advances in {{Neural Information
  Processing Systems}}}}, Vol.~\bibinfo{volume}{32}.
\newblock
\href{https://doi.org/10.5555/3454287.3455202}{doi:\nolinkurl{10.5555/3454287.3455202}}


\bibitem[Zhu and Tan(2023)]%
        {zhuvulaste2023}
\bibfield{author}{\bibinfo{person}{Botong Zhu} {and} \bibinfo{person}{Huobin
  Tan}.} \bibinfo{year}{2023}\natexlab{}.
\newblock \showarticletitle{{{VuLASTE}}: {{Long Sequence Model}} with
  {{Abstract Syntax Tree Embedding}} for {{Vulnerability Detection}}}. In
  \bibinfo{booktitle}{\emph{{{AAIA}} 2023 -- {{International Conference}} on
  {{Advances}} in {{Artificial Intelligence}} and {{Applications}}}}.
  \bibinfo{pages}{392--396}.
\newblock
\href{https://doi.org/10.1145/3603273.3635667}{doi:\nolinkurl{10.1145/3603273.3635667}}


\end{thebibliography}

\end{document}